\documentclass[sigconf, nonacm, pdfa]{acmart}

\usepackage{graphicx}
\usepackage{balance}  
\usepackage{epsfig}
\usepackage{graphicx}
\usepackage{epstopdf}
\usepackage{latexsym}
\usepackage{enumitem}
\PassOptionsToPackage{noend}{algpseudocode}
\usepackage{algpseudocode}
\usepackage{pifont}
\usepackage{epsfig}
\usepackage{amssymb}
\usepackage{amsmath}
\usepackage{amsfonts}
\usepackage{subfigure}
\usepackage{stmaryrd}
\usepackage{url}
\usepackage{multirow}
\usepackage{array}
\usepackage[normalem]{ulem}
\usepackage{color}
\usepackage{cleveref}
\usepackage{xspace}
\usepackage{mathtools}
\usepackage{soul}
\usepackage{listings}
\usepackage{enumitem}
\usepackage{xcolor}
\usepackage{tikz}
\newsavebox{\blackball}
\newsavebox{\greenball}

\usepackage[utf8]{inputenc}
\usepackage{enumitem}
\usepackage{amsmath}

\usepackage{bbding}
\usepackage{amssymb}
\usepackage{epstopdf}
\usepackage{epsfig,endnotes}
\usepackage{url}
\usepackage{array}
\usepackage{booktabs}
\usepackage{threeparttable}
\usepackage{tabulary}

\usepackage{CJKutf8}
\usepackage[most,breakable]{tcolorbox}
\usepackage{etoolbox}
\usepackage{fancyhdr}
\usepackage{lipsum}
\usepackage[fencedCode]{markdown}
\usepackage{tikz}

\usepackage[lined,boxed,vlined,ruled,linesnumbered]{algorithm2e}

\usepackage{bm}

\theoremstyle{definition}
\newtheorem{definition}{Definition}[section]

\newcommand{\revise}[1]{\textcolor{blue}{#1}}

\newcommand{\lgl}[1]{\textcolor{blue}{LGL: #1}}
\newcommand{\zxh}[1]{\textcolor{orange}{#1}}

\newcolumntype{M}[1]{>{\centering\arraybackslash}m{#1}}
\errorcontextlines\maxdimen

\newcommand{\oursys}{\textsc{D-Bot}\xspace}

\newcommand{\llm}{\textsc{LLM}\xspace}
\newcommand{\llms}{\textsc{LLMs}\xspace}

\newcommand{\languagemodels}{{large language models}\xspace}

\newcommand{\hi}[1]{\vspace{.25em} \noindent {\bf #1}\xspace}

 \newcommand{\bfit}[1]{\textbf{\textit{#1}}}

\newcommand{\human}{\emph{HumanDBA}\xspace}
\newcommand{\dbotfour}{\emph{D-Bot (GPT-4)}\xspace}
\newcommand{\dbotthree}{\emph{D-Bot (GPT-3.5)}\xspace}

\newcommand{\dnn}{\emph{DNN}\xspace}
\newcommand{\decisiontree}{\emph{DecisionTree}\xspace}

\newcommand{\vanillafour}{\emph{GPT-4}\xspace}
\newcommand{\vanillathree}{\emph{GPT-3.5}\xspace}
\newcommand{\noknow}{\emph{NoKnowledge}\xspace}
\newcommand{\noreflect}{\emph{NoTreeSearch}\xspace}
\newcommand{\single}{\emph{SingleLLM}\xspace}
\newcommand\vldbavailabilityurl{https://github.com/TsinghuaDatabaseGroup/DB-GPT}

\definecolor{codegreen}{rgb}{0,0.6,0}
\definecolor{codegray}{rgb}{0.5,0.5,0.5}
\definecolor{codepurple}{rgb}{0.58,0,0.82}
\definecolor{backcolour}{rgb}{0.95,0.95,0.92}

\lstdefinestyle{mystyle}{
	backgroundcolor=\color{backcolour},   
	commentstyle=\color{codegreen},
	keywordstyle=\color{magenta},
	numberstyle=\tiny\color{codegray},
	stringstyle=\color{codepurple},
	basicstyle=\ttfamily\footnotesize,
	breakatwhitespace=false,         
	breaklines=true,                 
	captionpos=b,                    
	keepspaces=true,                 
	numbers=left,                    
	numbersep=5pt,                  
	showspaces=false,                
	showstringspaces=false,
	showtabs=false,                  
	tabsize=2
}
\lstdefinestyle{jsonStyle}{
	basicstyle=\small\ttfamily,
	columns=fullflexible,
	showstringspaces=false,
	commentstyle=\color{codegreen}\upshape,
	stringstyle=\color{codegreen},
	morestring=[b]",
	moredelim=[s][\color{codepurple}]{\{}{\}},
	moredelim=[s][\color{codepurple}]{[}{]},
	moredelim=[l][\color{codepurple}]{:},
	moredelim=[l][\color{codepurple}]{,}
}

\setcopyright{none}
\begin{document}

\title{LLM As Database Administrator}
\title{LLM As DBA}
\title{D-Bot: Database Diagnosis System using Large Language Models}

\author{Xuanhe Zhou$^{1}$, Guoliang Li$^{1}$, Zhaoyan Sun$^{1}$, Zhiyuan Liu$^{1}$, Weize Chen$^{1}$, Jianming Wu$^{1}$}
\author{Jiesi Liu$^{1}$, Ruohang Feng$^{2}$, Guoyang Zeng$^{3}$}
\affiliation{%
	\institution{$^{1}$Tsinghua University  $^{2}$Pigsty $^{3}$ModelBest}
	\city{}
	\country{}
}
\email{zhouxuan19@mails.tsinghua.edu.cn, liguoliang@tsinghua.edu.cn}

\pagestyle{plain}
\pagenumbering{arabic}


\begin{abstract}
	\begin{sloppypar}
		Database administrators (DBAs) play an important role in managing, maintaining and optimizing database systems. However, it is hard and tedious for DBAs to manage a large number of databases and give timely response ({waiting for hours is intolerable in many online cases}). In addition, existing empirical methods only support limited diagnosis scenarios, which are also labor-intensive to update the diagnosis rules for database version updates. Recently \languagemodels (\llms) have shown great potential in various fields. Thus, we propose \oursys, an \llm-based database diagnosis system that can automatically acquire knowledge from diagnosis documents, and generate reasonable and well-founded diagnosis report (i.e., identifying the root causes and solutions) within acceptable time (e.g., under 10 minutes compared to hours by a DBA). The techniques in \oursys include $(i)$ offline knowledge extraction from documents, $(ii)$ automatic prompt generation (e.g., knowledge matching, tool retrieval), $(iii)$ root cause analysis using tree search algorithm, and $(iv)$ collaborative mechanism for complex anomalies with multiple root causes. We verify  \oursys on real benchmarks (including 539 anomalies of six typical applications),  and the results show that \oursys can effectively analyze the root causes of unseen anomalies and {\it significantly outperforms traditional methods and vanilla models like GPT-4}. 
	\end{sloppypar}
\end{abstract}

\maketitle

\ifdefempty{\vldbavailabilityurl}{}{
	\begingroup\small\noindent\raggedright\textbf{PVLDB Artifact Availability:}\\
	The source code, data, technical report, and other artifacts have been made available at \url{\vldbavailabilityurl}.
	\endgroup
}


\section{Introduction}
\label{sec:intro}

\begin{figure}[!t]
	\vspace{.5em}
	\centering
	\includegraphics[width=1.02\linewidth, trim={.5cm .25cm 0cm 0},clip]{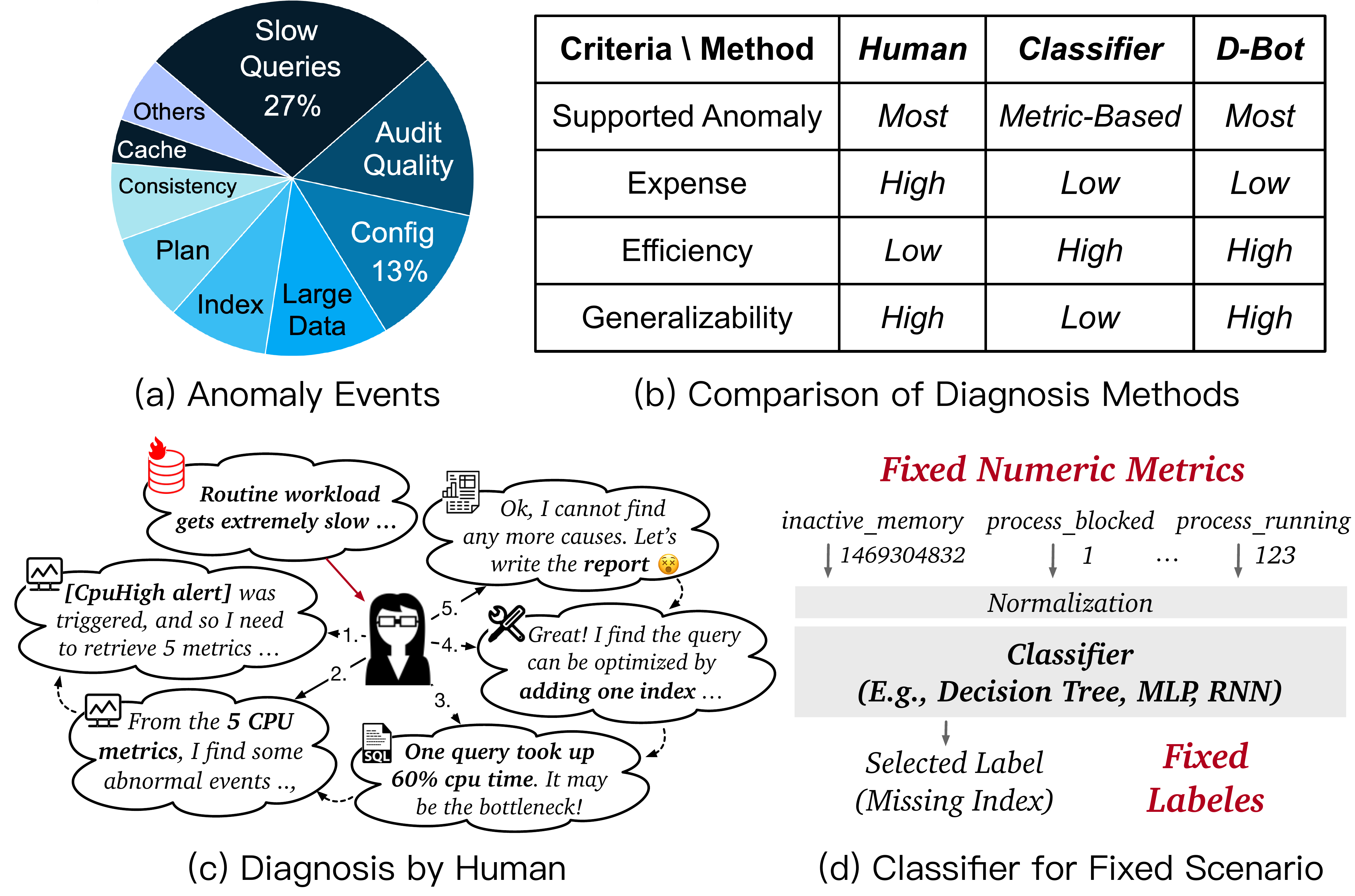}
	\vspace{.5em}
	\captionsetup{skip=4pt}
	\vspace{-1em}
	\caption{Database diagnosis is a complex problem mainly handled by human expertise -- (a) example root causes of anomalies in a database vendor; (b) comparison of diagnosis methods; (c) toy example of diagnosing by human DBA; (d) example of diagnosing by machine learning classifier.}
	\label{fig:intro}
	\vspace{-1em}
\end{figure}

Database diagnosis aims to detect, analyze, and resolve anomaly events in database systems, thereby ensuring high data availability and workload performance. However, database anomalies are remarkably diverse, making it impossible to comprehensively cover them with predefined rules~\cite{hai2021systematic}.  As shown in Figure~\ref{fig:intro} (a), a database vendor encountered over 900 anomaly events in three months, most of which spanned various facets of database and system modules (e.g., slow query processing, locking mechanisms, improper configurations). Furthermore, these modules exhibit complex correlations with system metrics (e.g., high CPU usage may result from concurrent commits or massive calculations). So it requires to explore different reasoning strategies (e.g., investigating different system views) before identifying the potential root causes.

As a result, database diagnosis is a challenging problem, where {\it ``the devil is in the details''}~\cite{DBLP:conf/sigmod/YoonNM16,DBLP:conf/cidr/DiasRSVW05}. Many companies rely on the expertise of human database administrators (DBAs) to undertake diagnosis tasks. Here we present a simplified example (Figure~\ref{fig:intro} (c)): {\it (1) Anomaly Notification.} The database user notifies an anomaly, e.g., ``routine queries $...$ is 120\% slower than the norm $...$''; {\it (2) Alert Detection.} Upon receiving
 the user's notification, the DBA first investigates the triggered alerts. For instance, the DBA discovers a ``CPU High'' alert, indicating the total CPU usage exceeded 90\% for 2 minutes; {\it (3) Metric Analysis.} Next the DBA delves deeper to explore more CPU-related metrics (e.g., the number of running or blocked processes, the number of query calls). By analyzing these metrics, the DBA concludes the issue was caused by some resource-intensive queries. {\it (4) Event Analysis.} The DBA retrieves the statistics of top-k slow queries (query templates) from database views, and finds one query consumed nearly 60\% of the CPU time. {\it (5) Optimization Advice.} The DBA tries to optimize the problematic query (e.g., index update, SQL rewrite) by experience or tools. 

The above diagnosis process is inherently iterative (e.g., if the DBA fails to find any abnormal queries, she may turn to investigate I/O metrics). Besides, the DBA needs to write a diagnosis report\footnote{Over 100 diagnosis reports are available on the website  \textit{\href{http://dbgpt.dbmind.cn/}{http://dbgpt.dbmind.cn/}}.} to facilitate the user's understanding, which includes information like root causes together with the detailed diagnosis processes.



However, there exists a {\it significant gap between the limited capabilities of human DBAs and the daunting diagnosis issues}. Firstly, training a human DBA demands an extensive amount of time, often ranging from months to years, by understanding a large scale of relevant documents (e.g., database tuning guides) and the necessity for hands-on practice. Secondly, it is nearly impossible to employ sufficient number of human DBAs to manage a vast array of database instances (e.g.  millions of instance on cloud databases). Thirdly, a human DBA may not provide timely responses in urgent scenarios, especially when dealing with correlated issues across multiple database modules, which potentially lead to significant financial losses. Thus, \textit{if typical anomalies can be automatically resolved, it will relieve the burden of human DBAs and save resources.}


\begin{sloppypar}
Driven by this motivation, many database products are equipped with semi-automatic diagnosis tools~\cite{DBLP:journals/pacmmod/HuangWZTL023,DBLP:conf/ccgrid/LuXL0NZYZSZP22,DBLP:conf/ipccc/LiuZSMYP20,DBLP:conf/sigmod/KalmeghBR19,DBLP:conf/icde/LiuYZGCGLWLTL22}. However, they have several limitations. First, they are built by empirical rules~\cite{DBLP:conf/sigmod/YoonNM16,DBLP:conf/cidr/DiasRSVW05} or small-scale ML models (e.g., classifiers~\cite{DBLP:journals/pvldb/MaYZWZJHLLQLCP20}), which have poor scenario understanding capability and cannot utilize the diagnosis knowledge. 
Second, they cannot be flexibly generalized to scenario changes. For empirical methods, it is tedious to manually update and verify rules by newest versions of documents. And learned methods (e.g., XGBoost~\cite{chen2016xgboost}, KNN~\cite{guo2003knn}) require to redesign the input metrics and labels, and retrain models for a new scenario (Figure~\ref{fig:intro} (d)). Third, these methods have no inference ability as human DBAs, such as recursively exploring system views based on the initial analysis results to infer the root cause. 
\end{sloppypar}

To this end, we aim to build an \emph{intelligent diagnosis system} with three main advantages. \bfit{(1) Precise Diagnosis.} First, our system can utilize tools to gather scenario information (e.g., query analysis with flame graph) or derive optimization advice (e.g., index selection),  which are necessary for real-world diagnosis. However, that is hardly supported by traditional methods. Second, it can conduct basic logical reasoning (i.e., making diagnosis plans).  \bfit{(2) Expense and Time Saving.} The system can relieve human DBAs from on-call duties to some extent (e.g., resolving typical anomalies that rules cannot support). \bfit{(3) High Generalizability.} The system exhibits flexibility in analyzing  unseen anomalies based on both the given documents (e.g., new metrics, views, logs) and past experience.

Recent advances in Large Language Models (\llms) offer the potential to achieve this goal, which have demonstrated superiority in natural language understanding and programming~\cite{zhou2022large,zheng2023codegeex,qin2023tool,DBLP:journals/corr/abs-2307-16789}. 
However, database diagnosis requires extensive domain-specific skills and {\it even the GPT-4 model cannot directly master the diagnosis knowledge ({lower than 50\% accuracy})}. This poses three challenges.


\bfit{(C1) How to enhance \llm's understanding of the diagnosis problem?} Despite pre-trained on extensive corpora, \llms still struggle in effectively diagnosing without proper prompting\footnote{Prompting is to add additional information into \llm input. Although \llms can memorize new knowledge with fine-tuning, it may forget previous knowledge or generate inaccurate or mixed-up responses, which is unacceptable in database diagnosis.} (e.g., unaware of the database knowledge). The challenges include  $(i)$ extracting useful knowledge from long documents (e.g., correlations across chapters); $(ii)$ matching with suitable knowledge by the given context (e.g., detecting an alert of high node load); $(iii)$ retrieving tools that are potentially useful (e.g., database catalogs).




\begin{sloppypar}
\bfit{(C2) How to improve \llm's diagnosis performance for single-cause anomalies?} With knowledge-and-tool prompt, \llm needs to judiciously reason about the given anomalies. First, different from many \llm tasks~\cite{ding2021openprompt}, database diagnosis is an interactive procedure that generally requires to analyze for many times, while \llm has the early stop problem~\cite{dodge2020fine}. Second, \llm has a ``hallucination'' problem~\cite{rawte2023survey}, and it is critical to design strategies that guide \llm to derive in-depth and reasonable analysis. 
\end{sloppypar}


\bfit{(C3) How to enhance \llm's diagnosis capability for multi-cause anomalies?} 
From our observation, within time budget, a single \llm is hard to accurately analyze for complex anomalies (e.g., with multiple root causes and the critical metrics are in finer-granularity). Therefore, it is vital to design an efficient diagnosis mechanism where multiple \llms can collaboratively tackle complex database problems (e.g., with cross reviews) and improve both the diagnosis accuracy and efficiency.





To tackle above challenges, we propose \oursys, a database diagnosis system using large language models. First, we extract useful knowledge chunks from documents (summary-tree based knowledge extraction) and construct a hierarchy of tools with detailed usage instructions, based on which we initialize the prompt template for \llm diagnosis (see Figure~\ref{fig:overview}). Second, according to the prompt template, we generate new prompt by matching with most relevant knowledge (key metric searching) and tools (fine-tuned SentenceBert), which \llm can utilize to acquire monitoring and optimization results for reasonable diagnosis. 
Third, we introduce a tree-based search strategy that guides the \llm to reflect over past diagnosis attempts and choose the most promising one, which significantly improves the diagnosis performance. 
Lastly, for complex anomalies (e.g., with multiple root causes), we propose a collaborative diagnosis mechanism where multiple \llm experts can diagnose in an asynchronous style (e.g., sharing analysis results, conducting cross reviews) to resolve the given anomaly.

\hi{Contributions.} We make the following contributions.

\noindent (1) We design an \llm-based database diagnosis framework to achieve precise diagnosis  (see Section~\ref{sec:overview}).

\noindent (2) We propose a diagnosis prompt generation method that empowers \llm to perform diagnosis by $(i)$ matching with relevant knowledge extracted from documents and $(ii)$ retrieving tools with a fine-tuned embedding model (see Sections~\ref{sec:docs} and \ref{sec: prompt}).

\noindent (3) We propose a root cause analysis method that improves the diagnosis performance using tree-search-based algorithm that guides \llm to conduct multi-step analysis (see Section~\ref{sec:diag}).

\noindent (4) We propose a collaborative diagnosis mechanism to improve the diagnosis efficiency, which involves multiple \llms concurrently analyzing issues by their domain knowledge (see Section~\ref{sec:collaborate}). 

\noindent (5) Our experimental results demonstrate that \oursys can accurately identify typical root causes within acceptable time (see Section~\ref{sec:experiments}).

\begin{figure*}[!t]
	\centering
	\includegraphics[width=1.\linewidth, trim={0 0 0 0},clip]{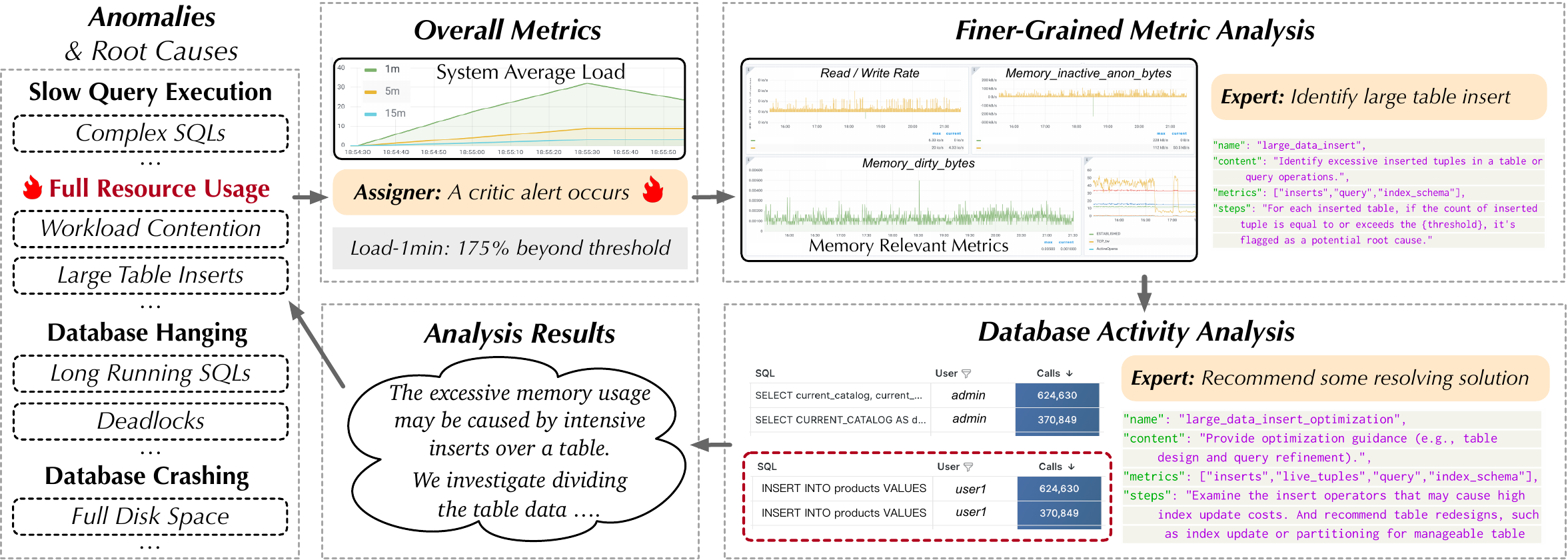}
	\vspace{-2em}
	\caption{Example of Database Diagnosis.}
	\label{fig:observe}
	\vspace{-1.25em}
\end{figure*}

\section{Preliminaries}
\label{sec:preliminary}

\subsection{Database Anomalies}

\emph{Anomalies} refer to the irregular or unexpected issues that necessitate the diagnosis procedure~\cite{ramachandran2018anomaly}. Figure~\ref{fig:observe} show four typical anomalies in databases.

\textit{(1) Slow Query Execution.} The database experiences longer response time than expectancy. For example, the slow query causes significant increase in CPU usage (system load) and query duration time, but the number of active processes remains low. 

\textit{(2) Full Resource Usage.} Some system resource is exhausted, preventing it accepting new requests or even causing errors (e.g., insert failures for running out of memory). For example, the high concurrency workload can not only cause great CPU and memory usage, but significantly increases the number of active processes. 

\textit{(3) Database Hanging.} The database becomes unresponsive, which is usually caused by long-running queries, deadlocks, or resource contention. For example, abnormal waits in submitted transactions cause great CPU consumption like the {\it full resource usage} anomaly, but it also involves frequent process interrupt and switching. 

\textit{(4) Database Crashing.} The database unexpectedly shuts down, causing data to become inaccessible. A typical root cause is full disk space, which leads to an inability to write new data or perform necessary operations, ultimately resulting in database failure and releasing the acquired resources.

\subsection{Database Diagnosis}
\label{subsec:def-diag}

\emph{Database diagnosis} refers to the process of analyzing and resolving above anomalies (usually in the form of a series of alerts) that occur within the database system. The primary objective of database diagnosis is to pinpoint the underlying \emph{root causes}. Here we showcase some example root causes in standalone databases\footnote{Anomalies on the application or network side fall outside the scope of this work.}:

\textit{(1) Concurrency Workloads}: Problems characterized by severe workload contention, where multiple database operations compete for system resources, leading to performance degradation.

\textit{(2) Query Operator Issues}: Problems like inserting large tables, fetching large volumes of data, and executing complex predicates, which can strain the database system's processing capabilities.

\textit{(3) Planning and Execution}: Root causes in this category involve abnormal planning times and prolonged database wait times, indicating inefficiencies in query planning and execution processes.

\textit{(4) Data-specific Issues}: Problems like data corruption and dead tuples (rows that are no longer needed but remain in the physical storage) may lead to performance problem.

\textit{(5) Database Schema and Settings}: These issues related to database schema (e.g., indexes) and configuration settings. Examples include missing indexes and small shared buffer sizes, which can impact query optimization and memory management.


\textit{(6) Harmful Background Tasks}: Some database maintenance tasks, like ``vacuum'' for storage space reclamation, can become problematic when invoked too frequently (these tasks will compete system resources with user queries).




Once the root causes are identified, a set of \emph{optimization actions} can be proposed to resolve these issues and restore normal database operations. Here we showcase some optimization tools.

\textit{(1) Query Rewrite Tools.} Since most databases are weak in logical transformations~\cite{wang2022wetune} (e.g., complex predicate simplification), there are external rewrite tools (e.g., around 120 rules in Calcite~\cite{begoli2018apache}) that help to optimize slow queries. 

\begin{sloppypar}
\textit{(2) Knob Tuning Tools.} Improper knob values may cause database failures (e.g., exceeding the maximal connection number) or bad performance (e.g., allocated working memory is too small). Thus, there are tools that utilize rules to provide tuning suggestions~\cite{pgtune,DBLP:conf/sigmod/Trummer22}. For instance, it increases the value of innodb\_buffer\_pool\_size in MySQL by 5\% if the memory usage is lower than 60\%. 
\end{sloppypar}

\textit{(3) Index Tuning Tools.} Similarly, there are index tuning rules that generate potentially useful indexes~\cite{DBLP:conf/cikm/LanBP20,AutoAdmin,Drop,DB2Advis,zhou2022autoindex}, such as creating composite index with columns in the same predicate. 







\begin{example}
	\begin{sloppypar}
		{\it As shown in Figure~\ref{fig:observe}, given an anomaly alert indicating high memory usage, we first examine the system load (e.g., node\_memory\_total for memory usage) during the anomaly time. The data confirms an abnormal memory utilization (over 90\%). To understand this, we further obtain the relevant memory metrics (e.g., node\_memory\_inactive\_anon\_bytes). Analysis of these metrics suggests that the excessive memory usage may be caused by an intensive workload that inserts data into a table. To address this, we investigate if optimization strategies could help with reducing the memory consumption (e.g., dividing table data into partitions).}
	\end{sloppypar}
\end{example}

\begin{figure*}[!t]
	\centering
	\includegraphics[width=1\linewidth, trim={0 0em 0cm 0},clip]{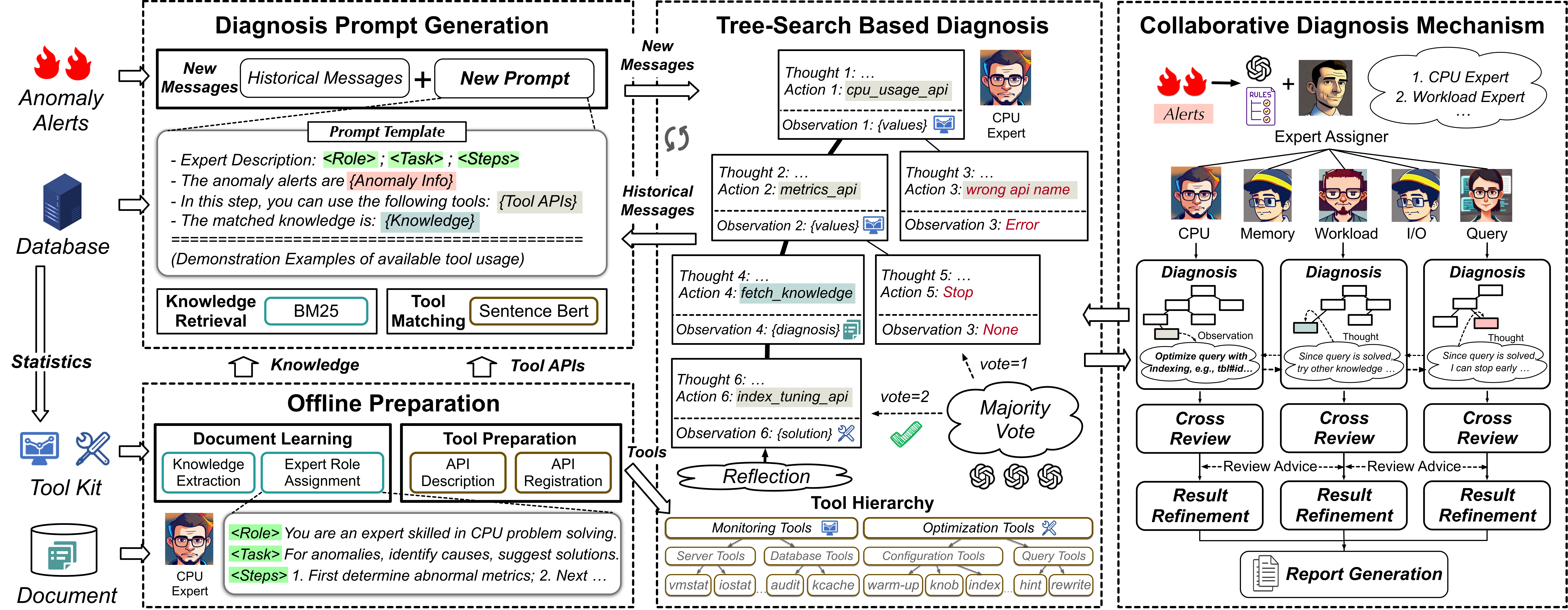}
	\vspace{-2em}
	\caption{Database Diagnosis in \oursys.}
	\label{fig:overview}
	\vspace{-1em}
\end{figure*}

\subsection{Large Language Models}
\label{subsec:llm}

Next, we introduce the fundamental concepts of Large Language Models (LLMs), including \llm architecture, \llm Prompting, and \llm Fine-tuning, which are pivotal for harnessing their capabilities in database diagnosis.

\begin{sloppypar}
\bfit{Transformer-Based \llms.} Existing \llms mainly adopt the Transformer architecture, distinguished by its attention mechanism and feed-forward neural networks. Attention mechanism dynamically weighs elements in the input, allowing the model to focus on different parts of the text, i.e., the attention scores are computed as ${Attention}(Q,K,V) = \text{softmax}\left(\frac{QK^T}{\sqrt{d_k}}\right)V$, where $Q$ (queries), $K$ (keys), and $V$ (values) represent different aspects of the text, and $d_k$ is the dimension of the keys. In addition, \llms include feed-forward neural networks in each layer, which apply position-wise linear transformations to the output of the attention layer. This combination of attention and feed-forward networks facilitate accurate predictions of subsequent text elements.	
\end{sloppypar}

\bfit{\llm Prompting.} To provide \llm with specific instructions for guiding their response generation, we can prepend or append a prompt \( P \) to the input \( X \) to create a new input, denoted as  \(  X' = P \oplus X \) or \(X' =  X \oplus P \). \llm then generates the output text based on this modified input. Note $(i)$ prompting does not require any additional updates to the model parameters and $(ii)$ prompts can be manually crafted or automatically learned from data~\cite{zhou2022large}.

\bfit{LLM Fine-tuning} involves adjusting the model parameters on a small and task-specific dataset (e.g., thousands of samples). Initially, the model parameters, denoted as \(\theta\), are inherited from the pre-training phase. Fine-tuning aims to minimize the loss function \(\mathcal{L}\), tailored to the specific task (e.g., classification or regression), over the task-specific dataset \(\mathcal{D}\). It is represented as
\(\theta_{\text{new}} = \theta_{\text{old}} - \alpha \cdot \nabla \mathcal{L}(\theta_{\text{old}}; \mathcal{D}),\)
where a small learning rate \(\alpha\) is often used to ensure gradual parameter updates~\cite{zeng2023openbmb}.

We rely on \llm prompting to guide close-sourced \llms like GPT-4 to diagnose (see Section~\ref{sec: prompt}), and utilize \llm fine-tuning to prepare localized \llms (see Section~\ref{subsec:finetune}).

\section{The Overview of \oursys}
\label{sec:overview}

\begin{figure}[!t]
	\centering
	\includegraphics[width=1\linewidth, trim={0 0em 0cm 0},clip]{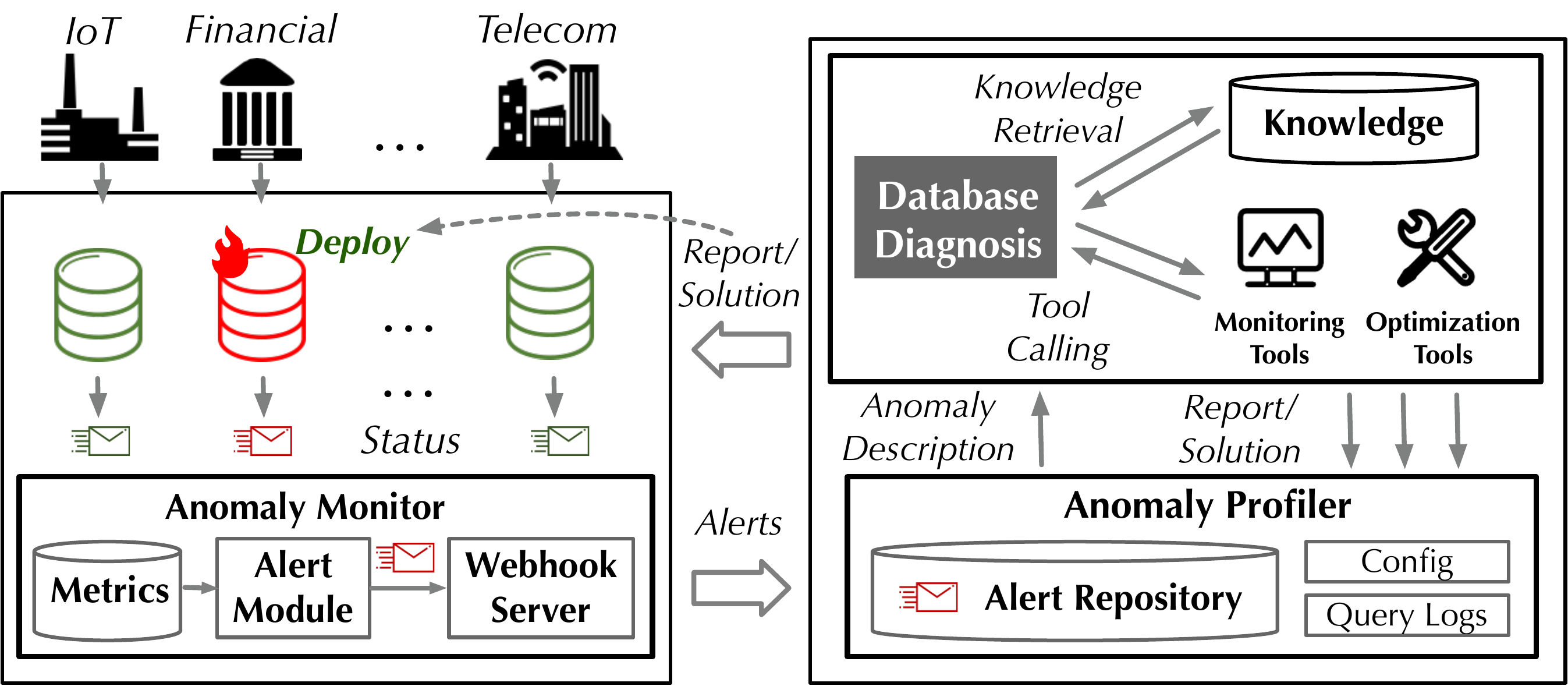}
	\vspace{-1.75em}
	\caption{\oursys Architecture.}
	\label{fig:arch}
	\vspace{-2.25em}
\end{figure}

\hi{Architecture.}  
As shown in Figure~\ref{fig:arch}, we first illustrate the overall architecture of \oursys.  (1) \emph{Anomaly Monitor} continuously monitors the status of databases using alert rules. (2) When a database turns to abnormal status, \emph{Anomaly Monitor} sends the triggered alerts to \emph{Anomaly Profiler}. (3) \emph{Anomaly Profiler} generates an anomaly description based on both the alerts and basic database information (e.g., query statistics during the time period of the alerts). (4) \emph{Database Diagnosis} empowers one or multiple \llm experts to collaboratively generate an analysis report for the anomaly. 

Next, we present the challenges and techniques in the database diagnosis component\footnote{This paper is an extension of the vision~\cite{zhou2023llm}} (see Figure~\ref{fig:overview}).

\bfit{(1) Offline Preparation.} \emph{Offline Preparation} equips \oursys with the knowledge and tools necessary for database diagnosis. 

$(i)$ \emph{Document Learning:} As diagnosis documents are not commonly found in the pre-training corpus and document sections have relations like cross-reference, we conduct knowledge extraction that builds summary trees to represent the document structures and extract knowledge chunks from the trees (e.g., traversing parent-child nodes and nodes with similar summaries). 

$(ii)$ \emph{Tool Preparation:} This step configures the diagnosis tools, such as providing API descriptions (explaining what API does, how it's used, and what kind of data it returns) and registering APIs into the system so they can be used by \llm during diagnosis. 

We cluster all the extracted knowledge chunks (Figure~\ref{fig:cluster}) and, for each cluster, we generate a prompt template (e.g., expert role, task description, basic diagnosis steps, available tools) to represent the characters of an \llm expert. For example, one \llm expert is in charge of analyzing CPU problems (knowledge) and can leverage tools like {\it cpu\_usage\_api} and {\it index\_tuning\_api}. In this way, each \llm expert handles a specific area of database problems.

\begin{figure}[!t]
	\vspace{.5em}
	\centering
	\includegraphics[width=1\linewidth, trim={0 0 0 0},clip]{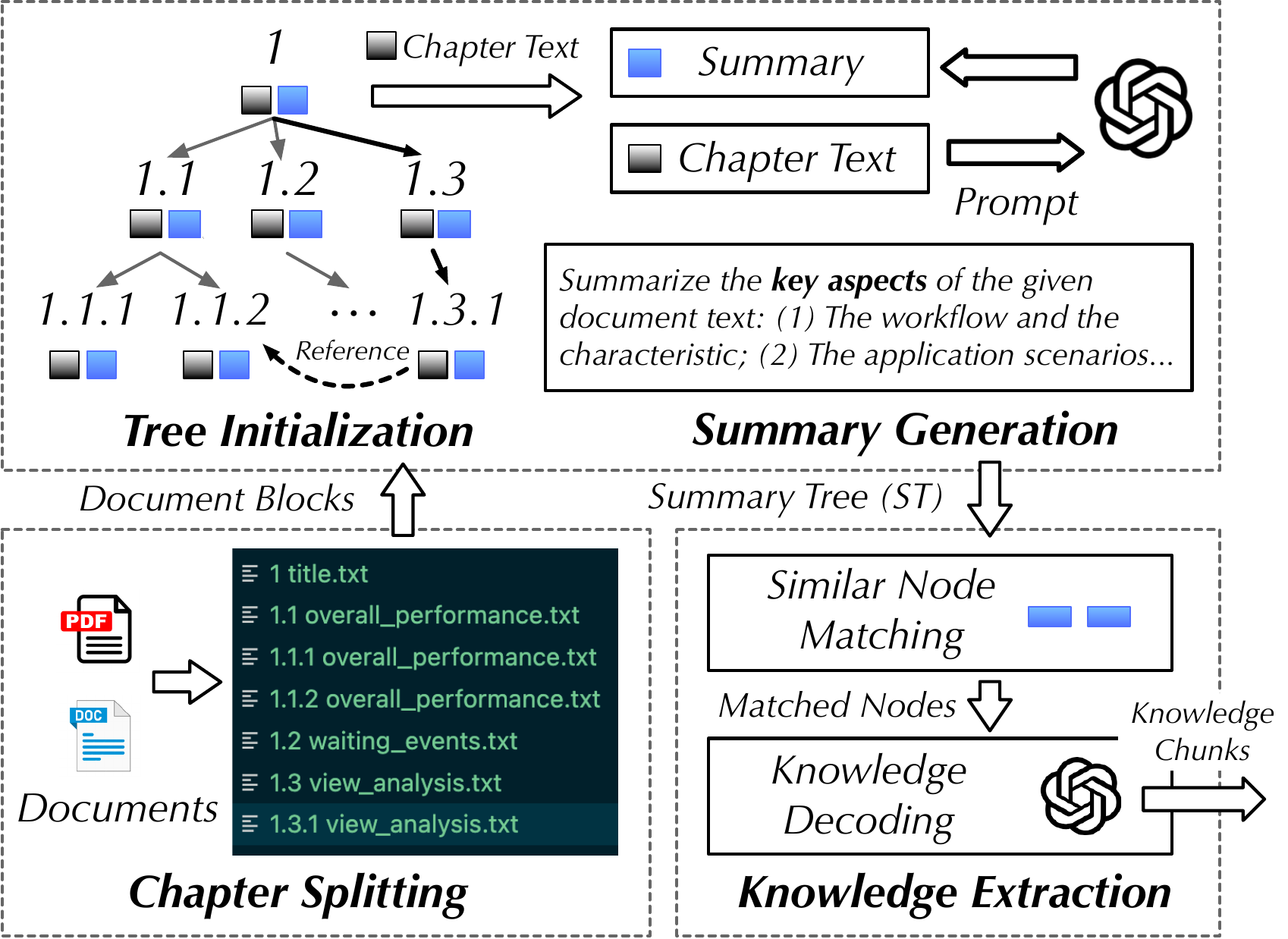}
	\vspace{-1em}
	\caption{Document Learning.}
	\label{fig:example}
	\vspace{-1.5em}
\end{figure}

\bfit{(2) Diagnosis Prompt Generation.} With all necessary knowledge and tools readily available, \emph{Diagnosis Prompt Generation} is responsible for creating context-aware prompts that guide the diagnosis process. Following the prompt template, each new prompt in the actual diagnosis requires three essential types of information: 

$(i)$ \emph{Anomaly Description.} The triggered alerts, including the occurring time, a summary of anomaly, severity level (e.g., warning or critical), additional features (e.g., anomaly status like ``resolved'').

$(ii)$ \emph{Tools}. 
Due to the variety of tools for different database usage (e.g., monitoring, indexing, query rewrite, query hint), we fine-tune a pre-trained Sentence-BERT model~\cite{DBLP:conf/emnlp/ReimersG19} to match to the current context (e.g., when \llm is unaware of which metrics is abnormal) with relevant tools (e.g., the {\it fetch\_abnormal\_metrics} API). 

$(iii)$ \emph{Knowledge Chunks} that instruct \llm on how to diagnose with the obtained abnormal metrics. We identify the most relevant knowledge chunks with keyword search methods (e.g., BM25). 

$(iiv)$ \emph{Historical Messages} containing valuable information, such as the past tool calling results, which are crucial for tree-search-based diagnosis (e.g., making decisions by previous steps).


\bfit{(3) Tree-Search Based Diagnosis.} \emph{Tree-Search Based Diagnosis} aims to identify the potential root causes and solutions using the generated prompt (calling tools or analyzing by following the knowledge).  For database diagnosis, \llm encounters challenges like hallucination and unstable \llm responses (e.g., incorrect API requests, overly general analysis) that can cause diagnosis failures. To solve this problem, we employ the {\it tree of thought} strategy, where \llm explores multiple possible reasoning chains and selects the most beneficial chain to explore (based on both the \llm's votes and selection frequency). Additionally, through the {\it reflection} mechanism, \llm can backtrack to {\it previous steps} if the current step fails to detect useful information, which significantly increases the likelihood of arriving at reasonable diagnosis results.


\bfit{(4) Collaborative Diagnosis Mechanism.} Since the cost of \emph{Tree-Search Based Diagnosis} increases significantly with the number of equipped knowledge and tools, \emph{Collaborative Diagnosis Mechanism} aims to improve the diagnosis performance by  leveraging multiple \llm experts. Given an anomaly, we begin by selecting relevant experts based on the triggered alerts and anomaly complexity. Next, the selected experts separately analyze ({\it Tree-Search Based Diagnosis}) with {\it a more focused set of tools and knowledge chunks}. Here we adopt an asynchronous strategy: the selected experts share intermediate diagnosis results during diagnosis, facilitating real-time information exchange and reducing the redundant analysis (e.g., the same query issues). After diagnosis, these experts conduct cross-review (e.g., identifying overly general root causes), based on which they iteratively refine the analysis results and ultimately generate a comprehensive report under specific template (e.g., including the background, root causes, solutions, and detailed steps).

\section{Offline Preparation}
\label{sec:docs}

In this section, we explain how to prepare necessary knowledge and tools for \llm diagnosis.




\subsection{Document Learning}

We first decide the knowledge format that is suitable to use in \llm prompting. Next, we introduce the extraction method to obtain such knowledge chunks from given documents. Finally, we showcase the obtained knowledge chunks and their clustering results.

\subsubsection{Knowledge Format}

Similarly to the diagnosis evidence in Figure~\ref{fig:observe}, given some documents, the desired knowledge chunk is composed of four parts: {\it (i) ``Name''} helps \llm to understand the overall function; {\it (ii) ``Content''} explains how the root cause can impact the database performance (e.g.,  performance degradation due to an excessive number of dead tuples); {\it (iii) ``Metrics''} is a list of involved metric names, used for knowledge matching in prompt generation (Section~\ref{subsec:cases}); {\it (iv) ``Steps''} provides the detailed procedure of analyzing with the relevant metrics. This allows the \llm to imitate and perform step-by-step analysis.

\subsubsection{Knowledge Extraction}
Next we explain how to extract such knowledge from documents. In database diagnosis, the relevant documents have two characters, i.e., $(i)$ most documents are of long context involved diversified aspects (e.g., both resource and configuration issues are discussed in maintenance guide) and $(ii)$ some paragraphs are correlated with each other. For example, the concept of ``bloat-table'' appearing in ``many\_dead\_tuples'' (like Chapter 3.2) is explained in another section (like Chapter 1.1.3). 

Although there are already some long-context \llms~\cite{openai} that support long documents as input, they cannot ensure the quality of answered knowledge (e.g., missing or making up important details~\cite{li2023loogle,xiong2023effective}). Thus, we propose a deterministic knowledge extraction algorithm in three steps.




\textit{Step1: Chapter Splitting.} Instead of directly splitting documents into fixed-length segments, we divide them based on the chapter structures and their content (e.g., applications split by keywords like ``tenant examples''). If a block exceeds the maximum block size (e.g., 4k tokens) that the \llm can handle, we further divide it recursively into smaller blocks.

\textit{Step2: Summary Tree Construction.} Next, based on the chapter relations, we initialize a tree structure, where the root node is the document title and other nodes denote split document blocks. For each node $i$, its child node denotes a subsection of chapter $i$ and node $i$ includes two parts: $(1)$ the content of chapter $i$ and $(2)$ the summary of chapter $i$, which is created by feeding the content into \llm with a summarization prompt, i.e., \definecolor{deepgray}{gray}{0.1}
\textcolor{deepgray}{\bfit{$p_{summarize}$} = {\it Summarize the provided chunk briefly $\cdots$ Your summary will serve as an index for others to find technical details related to database maintenance $\cdots$ Pay attention to examples even if the chunks cover other topics.}}

\begin{figure}[!t]
	\vspace{.5em}
	\centering
	\includegraphics[width=.9\linewidth, trim={22em 5em 10em 5em},clip]{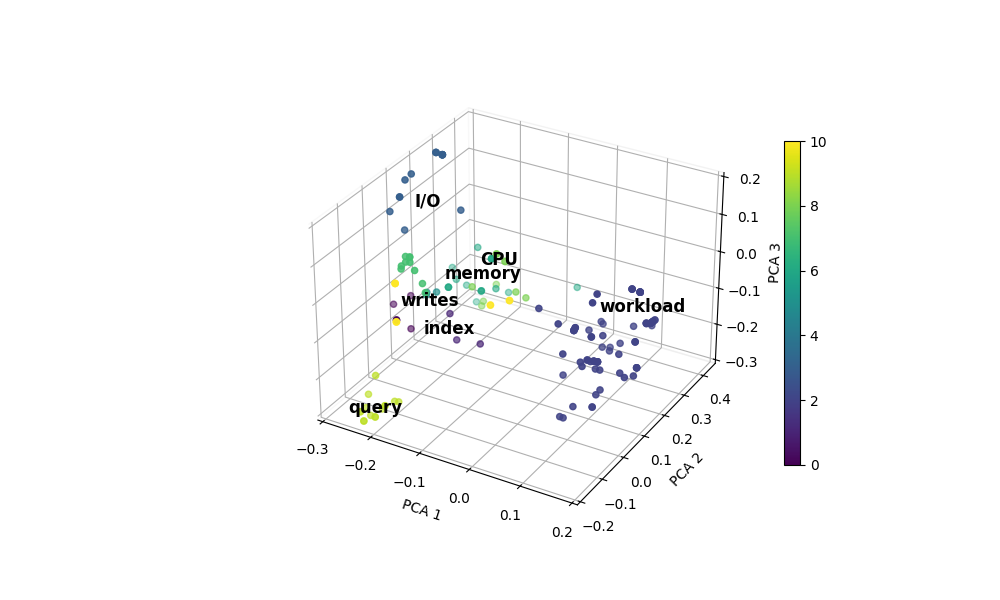}
	\vspace{-1.em}
	\caption{Clustering results of extracted knowledge.}
	\label{fig:cluster}
\end{figure}

The generated summary acts as a textual index of the node $i$, enabling the matching of blocks with similar content or relations like cross references.

\textit{Step3: Knowledge Extraction.} After generating the summary tree, \llm parses each document block $i$ (with content from both node $i$ and its child nodes) and compares it with the summaries of other blocks having similar content, which is guided by the extraction prompt, i.e., \definecolor{deepgray}{gray}{0.1}
\textcolor{deepgray}{\bfit{$p_{extract}$} = {\it ``Given a chunk summary, extract diagnosis experience from the chunk.} {\it If uncertain, explore diagnosis experience in chunks from child nodes or chunks with similar summaries.''}} 

This way, knowledge that correlates with the key points from the summaries are detected. For each detected knowledge $C_i$, we decide whether to keep $C_i$ in a hybrid manner. Specifically, if \llm indicates a low likelihood that $C_i$ is redundant (compared with existing knowledge), we will incorporate it. Otherwise, we will conduct a manual examination of $C_i$, where $C_i$ can be kept if we discover any new insights, even though $C_i$ has significant overlap with some existing knowledge. In this way, we can ensure {\it the inclusion of most diagnosis knowledge and reduce the potential for redundant information}. 

\subsubsection{Clustering Results of Extracted Knowledge} 

\label{subsubsec: cluster}

We showcase 188 knowledge chunks extracted from 81 pages of documents, including the general diagnosis guides, cases, and detailed reports\footnote{\href{https://github.com/TsinghuaDatabaseGroup/DB-GPT/tree/main/doc2knowledge/docs}{github.com/TsinghuaDatabaseGroup/DB-GPT/tree/main/doc2knowledge/docs}}. 

To derive insights from this diverse set of knowledge chunks, we $(i)$ convert the chunks into numerical vectors using a pre-trained embedding model (e.g., Ada-002~\cite{openai}); and $(ii)$ apply the DBSCAN algorithm~\cite{khan2014dbscan} to group knowledge chunks by the similarity of their text embeddings; and $(iii)$ reduce the dimensionality of the text embeddings (to three dimensions) using Principal Component Analysis (PCA). In this way, we can visualize the knowledge extraction results in Figure~\ref{fig:cluster}, which illustrates that {\it the knowledge distribution largely aligns with the types of root causes} (Section~\ref{subsec:def-diag}). Here, we highlight some knowledge topics. \underline{\emph{(1) Workloads}} include $(i)$ diagnosing workload contention issues (e.g., excessive connections), $(ii)$ analyzing the impact of database wait events, $(iii)$ addressing abnormal persistent events, $(iv)$ tackling long/short-term performance fluctuations. \underline{\emph{(2) Query Operators}} include $(i)$ analyzing inaccurate operator costs that influence query plans, $(ii)$ identifying slow operators (e.g., poor joins, aggregations, index filtering), $(iii)$ analyzing the impact of abnormal SQL structures. \underline{\emph{(3) Index Issues}} include $(i)$ identifying unused or redundant indexes, $(ii)$ addressing the performance impact of too many indexes in a table, especially on insert and update operations, $(iii)$ detecting missing indexes, and $(iv)$ analyzing why an index was used in a query plan (e.g., index invalidation, implicit type conversion).

It is evident that a knowledge chunk can be relevant to multiple topics (e.g., slow queries may get involved in both CPU and operator analysis). Thus, effective utilization and communication of these knowledge chunks (e.g., experts from different topics) are vital for the following diagnosis.

\subsection{Tool Preparation}

Apart from knowledge, human DBAs need to frequently interact with monitoring and optimization tools (e.g., database views, system commands, index tuning tools). To facilitate effective \llm diagnosis, it's essential to ensure \llm understand the complex API functions within available tools.

First, we establish a structured hierarchy to classify and organize ``categories-tools-APIs'', where ``APIs'' represent the specific functions of a tool. For example, an index selection tool would be categorized under ``optimization'', with ``configuration tool'' as its tool type, and ``heuristic\_index\_selection'' as an example API (Figure~\ref{fig:overview}). This hierarchy aids in organizing and understanding the diverse range of database tools.

Second, for each tool function, we provide a detailed \emph{utilization specification} (in the form of function comment). This includes the function's explanation, its parameters, and relevant use cases (for Section~\ref{subsec:tool}). For instance, the function explanation for ``heuristic\_index\_selection'' could be {\it ``Automatically select cost-reduction indexes based on query patterns and workload. Arguments include query frequency, data volume, index storage constraints, ...''}.

\begin{sloppypar}
Finally, we dynamically register tool functions by iterating through APIs in the given tool modules, obtaining each API's function names along with their \emph{utilization specifications}.
\end{sloppypar}

\section{Diagnosis Prompt Generation}
\label{sec: prompt}

Next we explain how to automatically generate diagnosis prompts by matching with the extracted knowledge and tools.

\subsection{Knowledge Retrieval}

\label{subsec:cases}

\begin{figure}[!t]
	\vspace{.5em}
	\centering
	\includegraphics[width=1.05\linewidth, trim={0 0 0 0},clip]{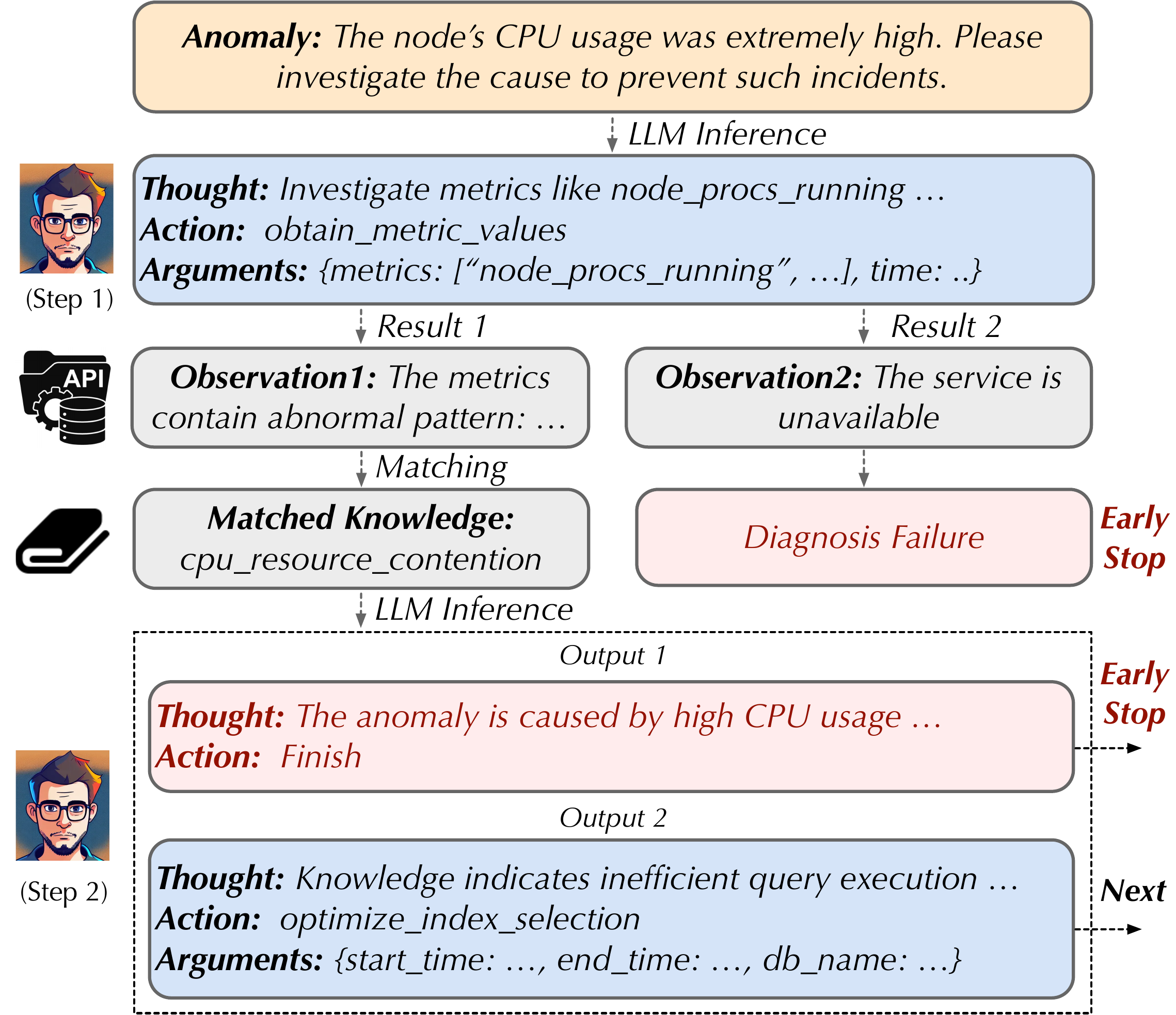}
	\vspace{-1.5em}
	\caption{Example multi-step diagnosis by \llm.}
	\label{fig:dignose}
	\vspace{-1em}
\end{figure}

Apart from knowledge that offers general diagnosis processes (included in the prompt template), most knowledge chunks are only useful under specific context. such as the analysis of abnormal CPU metrics (Figure~\ref{fig:dignose}). Thus, for a given context (e.g., with 5 abnormal CPU metrics), we adopt the approximate algorithm BM25~\cite{robertson2009probabilistic} to rank the most relevant knowledge chunks. Specifically, the BM25 algorithm ranks a set of knowledge chunks based on their ``metrics'' attribute, computed as:
\begin{equation}
\text{Score}(D,Q) = \sum_{i=1}^n \text{IDF}(q_i) \cdot \frac{f(q_i, D) \cdot (k_1 + 1)}{f(q_i, D) + k_1 \cdot (1 - b + b \cdot \frac{|D|}{\text{avgDL}})}
\end{equation}
where \(D\) is a knowledge block, \(Q\) is a set of abnormal metrics (by anomaly detection algorithms like KS-Test~\cite{berger2014kolmogorov}), \( f(q_i, D) \) is the frequency of the metric \( q_i \) in \( D \), \( \text{avgDL} \) is the average knowledge block length, \(k_1\) and \(b\) are free hyper-parameters. \( \text{IDF}(q_i) \) is the inverse document frequency of the metric \( q_i \), computed as:
\begin{equation}
\text{IDF}(q_i) = \ln \left( \frac{N - n(q_i) + 0.5}{n(q_i) + 0.5} + 1 \right)
\end{equation}
where \( N \) is the total number of extracted knowledge chunks, and \( n(q_i) \) is the number of documents containing metric \( q_i \). 

The advantage of this approach is that we can match knowledge chunks even when the names or meanings of the metrics involved are not exactly the same and easily apply the extract knowledge across different monitoring tools or even systems.

\subsection{Tool Matching}
\label{subsec:tool}



Different from {\it matching knowledge chunks with abnormal metrics}, database tools involve complex APIs and the API names may not be directly relevant to the context (e.g., APIs like {\it sort\_remove} for slow queries). In this way, the \emph{BM25} algorithm may have relatively high error rates. Thus, we propose to fine-tune a more powerful pre-trained Sentence-BERT model~\cite{reimers2019sentence} that accurately matches suitable tools for an diagnosis context. This procedure includes two main steps, i.e., model fine-tuning and tool matching.


$\bullet$ \emph{Sentence-BERT Fine-tuning:} Let \( S = \{s_1, s_2, \ldots, s_n\} \) denote the set of diagnosis contexts, and \( T = \{t_1, t_2, \ldots, t_m\} \) denote the set of database tools. We aim to fine-tune a pre-trained Sentence-BERT model to comprehend the relational context between anomalies and database tools. The fine-tuning process is performed with a labeled dataset \( D = \{(s_i, t_j, y_{ij})\}_{i=1, j=1}^{n, m} \), where \( y_{ij} \) is the label indicating the relevance of tool \( t_j \) for a diagnosis context \( s_i \). The objective function is computed by cross-entropy loss:
\begin{equation}
\mathcal{L} = -\sum_{i=1}^{n}\sum_{j=1}^{m} y_{ij} \log(p_{ij}) + (1-y_{ij}) \log(1-p_{ij}),
\end{equation}
where \( p_{ij} \) is the predicted probability that tool \( t_j \) is relevant for anomaly \( s_i \), obtained by passing the concatenated embeddings of \( s_i \) and \( t_j \) through a sigmoid function.

$\bullet$ \emph{Suitable Tool Matching:} After fine-tuning, the model is employed to match the appropriate database tools for a new diagnosis context \( s \). The matching score between diagnosis context \( s \) and database tool \( t_j \) is computed as the cosine similarity between their embeddings, i.e.,
\begin{equation}
\text{sim}(s, t_j) = \frac{\text{emb}(s) \cdot \text{emb}(t_j)}{||\text{emb}(s)||_2 ||\text{emb}(t_j)||_2},
\end{equation}
where \( \text{emb}(.) \) denotes the embedding function of the fine-tuned Sentence-BERT model. The set of recommended database tools \( \hat{T} \) for diagnosis context \( s \) is obtained by selecting the top-\( k \) tools with the highest matching scores:

\begin{equation}
\hat{T} = \text{arg top}_k \{\text{sim}(s, t_j)\}_{j=1}^{m}.
\end{equation}

Finally, the selected top-\( k \) tools are integrated into the prompt, including their names, function descriptions, and argument lists, based on which \llms can generate calling requests and obtain tool execution results to enhance root cause diagnosis.

\section{Tree Search For \llm Diagnosis}
\label{sec:diag}


As shown in Figure~\ref{fig:dignose}, \llms easily make mistakes like $(i)$ generating the wrong tool calling request or receiving request failures (e.g., temporarily service unavailable) and $(ii)$ stopping diagnosis early without carefully reflecting over the proposed root causes. 
To solve these problems, we propose to tree-search based algorithm that can guide \llm to go back to previous actions if the current action fails or no valuable root causes can be found (Algorithm~\ref{alg:mcts}). 

\bfit{Step1: Tree Initialization.} We initialize a tree structure where each node represents an action (e.g. tool calling or analysis based on matched knowledge), and the edges represent the flow from one action to another. 

\bfit{Step2: Simulate Execution.} This step kicks off the execution of simulations starting from the root node to a leaf node (i.e., the end of a complete diagnosis). It involves selecting nodes based on specific standard (e.g., detected abnormal metrics). Here we utilize the UCT (Upper Confidence Bound applied to Trees) function~\cite{ottens2017duct} to decide on the traversal path, i.e., 
\( UCT(n) = \frac{W(n)}{N(n)} + C \sqrt{\frac{2\ln N(p)}{N(n)}} \)
 , where $W(n)$ is the number of wins after the $n$-th move,
$N(n)$ is the= number of simulations after the $n$-th move,
$N(p)$ is the total number of simulations for parent node, and
$C$ is the exploration hyper-parameter. During the traversal, nodes with the highest UCT values are selected. 

Within the UCT function, $W(n)$ is computed by the voting of several evaluation \llms (e.g., three). For each leaf nodes, the \llms are presented with the scenario context and historical actions leading to the node. Based on this information, they cast votes in favor of the promising leaf nodes (e.g., based on the number and accuracy of involved root causes). The node receiving the highest number of votes ($W(n)$) is determined as the most favorable path. 

\bfit{Step3: Existing Node Reflection.} For each node in the path from the root node to the selected node, we depend on \llm to rethink the benefits of taking the action (e.g., prompting with {\it ``make some reflection to inherit to later trails''}), which are appended to the prompt of child node. It  can not only improve the analysis quality of next steps. Besides, if \llm decides there is no useful information, the node will be marked with {\it ``pruned''} so as to enhance the diagnosis efficiency.

\bfit{Step4: Terminal Condition.} If \llm cannot find any more root causes (leaf nodes) for a threshold time (e.g., 20 turns), the algorithm ends by outputting the root causes and solutions of the best node.


\begin{algorithm}[!t]
	\SetAlgoNlRelativeSize{-1}
	\SetNlSty{textbf}{(}{)}
	\SetAlgoNlRelativeSize{0}
	
	\KwData{$T$: Set of tool APIs; $K$: Set of knowledge chunks}
	
	Initialize tree\;
	
	\While{threshold time not reached}{
		\ForEach{node in tree}{
			Calculate $UCT$ for {\it node} \;
		}
		Select node with highest $UCT$\;
		Extend child nodes using tool API $t \in T$ or chunk $k \in K$\;
		\ForEach{action in extended nodes}{
			Reflect on potential benefits of the {\it action} with \llm\;
			Append reflection to the {\it node} \;
			\If{action adds no info}{Prune {\it node} \;}
		}
	}
	\Return{Results of leaf node with most votes}\;
	\caption{Tree-Search Based Diagnosis}
	\label{alg:mcts}
\end{algorithm}

\section{Collaborative Diagnosis For Complex Anomalies}
\label{sec:collaborate}

With tool leaning and tree search algorithm, the diagnosis accuracy of single \llm can be greatly improved. Nevertheless, we find single \llms have trouble in resolving complex anomalies with multiple root causes (e.g., looping over limited causes and struggling to identify additional ones).  To address this, we propose a collaborative mechanism where multiple \llms, each equipped with tools and tree search algorithms, work collectively to tackle complex cases~\cite{agentverse}.


\bfit{Step1: Expert Preparation.} We initialize 7 \llm experts by the knowledge clustering results (Section~\ref{sec:docs}). Each expert is equipped with different knowledge and necessary tools in the prompt. 

\bfit{Step2: Expert Assignment.} Next, to avoid resource waste and improve diagnosis efficiency, we assign appropriate experts to diagnose. That is, given an anomaly, we first generate a description of the anomaly (e.g., time period, alert types, severity level). Next, based on the anomaly description, \emph{Expert Assigner} utilizes an \llm (e.g., GPT-4) to select a set of most relevant experts. For example, \emph{CPU Expert} for the {\it Load\_High} alert and \emph{Memory Expert} for the {\it Out\_of\_Memory} alert. Note we adopt \llm rather than rules, which is more flexible to plugin new alert rules or expert roles.





\bfit{Step3: Asynchronous Diagnosis.} The chosen experts simultaneously diagnose (Section~\ref{sec:diag}). Despite utilizing a common \llm, each expert is uniquely equipped with role-specific settings and domain knowledge. We enhance the diagnosis process with an asynchronous communication mechanism~\cite{kreps2011kafka}, which is built on the publish-subscribe model. That is, experts ``publish'' their findings or updates, which are then automatically ``delivered'' to other experts who have ``subscribed'' to these specific types of updates (e.g., all the reset selected experts).  

This mechanism allows for the efficient and non-blocking information exchange (e.g., metric analysis, tool outputs, results) among \llm experts. For instance, the CPU Expert might post a finding about abnormal CPU load patterns of slow queries, triggering an event-driven notification to other experts. This event-driven approach enables the memory expert to promptly detect memory swap activities potentially caused by these slow queries.

\begin{sloppypar}
\bfit{Step4: Cross Review.} Although the experts own different domain knowledge (e.g., {\it os\_resource\_contention} for CPU Expert and {\it swap\_usage\_analysis} for Memory Expert), some common analysis evidence (e.g., operator types, configuration settings) of these experts may inspire each other. Thus, after obtaining the diagnosis results of all the experts, we conduct cross review and there are three main sub-steps:
\end{sloppypar}

\begin{table*}[]
	\caption{Micro Benchmark Statistics. The applications cover ten typical root causes (introduced in Section~\ref{subsec:def-diag}).}
	\label{tab:micro-bench}
	\vspace{-1em}
	\small
	\begin{tabular}{|l|l|l|l|l|l|l|l|l|l|l|l|}
		\hline
		\multicolumn{1}{|c|}{\textbf{\footnotesize{Application}}} & \multicolumn{1}{c|}{\textbf{\begin{tabular}[c]{@{}c@{}}\footnotesize{Sync}\\ \footnotesize{Commits}\end{tabular}}} & \multicolumn{1}{c|}{\textbf{\begin{tabular}[c]{@{}c@{}}\footnotesize{Many}\\ \footnotesize{Inserts}\end{tabular}}} & \multicolumn{1}{c|}{\textbf{\begin{tabular}[c]{@{}c@{}}\footnotesize{High}\\ \footnotesize{Updates}\end{tabular}}} & \multicolumn{1}{c|}{\textbf{\begin{tabular}[c]{@{}c@{}}\footnotesize{Many}\\ \footnotesize{Deletes}\end{tabular}}} & \multicolumn{1}{c|}{\textbf{\begin{tabular}[c]{@{}c@{}}\footnotesize{Index}\\\footnotesize{Missing}\end{tabular}}} & \multicolumn{1}{c|}{\textbf{\begin{tabular}[c]{@{}c@{}}\footnotesize{Redundant}\\ \footnotesize{Indexes}\end{tabular}}} & \multicolumn{1}{c|}{\textbf{\begin{tabular}[c]{@{}c@{}}\footnotesize{Large}\\\footnotesize{Data Insert}\end{tabular}}} & \multicolumn{1}{c|}{\textbf{\begin{tabular}[c]{@{}c@{}}\footnotesize{Large}\\\footnotesize{Data Fetch}\end{tabular}}} & \multicolumn{1}{c|}{\textbf{\begin{tabular}[c]{@{}c@{}}\footnotesize{Poor}\\\footnotesize{ Join}\end{tabular}}} & \multicolumn{1}{c|}{\textbf{\begin{tabular}[c]{@{}c@{}}\footnotesize{Correlated} \\ \footnotesize{Subquery}\end{tabular}}} & \multicolumn{1}{c|}{\textbf{\footnotesize{Cases}}} \\ \hline
		Internet of Things                                        & $\checkmark$                                                                                    & $\times$                                                                                     & $\times$                                                                                     & $\times$                                                                                     & $\times$                                                                                     & $\times$                                                                                        & $\times$                                                                                         & $\times$                                                                                        & $\times$                                                                                & $\times$                                                                                           & 83                                    \\ \hline
		E-Commerce                                 & $\times$                                                                                     & $\checkmark$                                                                                    & $\checkmark$                                                                                    & $\checkmark$                                                                                    & $\checkmark$                                                                                    & $\checkmark$                                                                                       & $\checkmark$                                                                                        & $\checkmark$                                                                                       & $\checkmark$                                                                               & $\checkmark$                                                                                          & 211                                    \\ \hline
		Financial                                  & $\checkmark$                                                                                    & $\checkmark$                                                                                    & $\times$                                                                                     & $\times$                                                                                     & $\times$                                                                                     & $\checkmark$                                                                                       & $\checkmark$                                                                                        & $\checkmark$                                                                                       & $\checkmark$                                                                               & $\times$                                                                                           & 31                                     \\ \hline
		Business Intel.                                         & $\times$                                                                                     & $\checkmark$                                                                                    & $\checkmark$                                                                                    & $\checkmark$                                                                                    & $\checkmark$                                                                                    & $\checkmark$                                                                                       & $\checkmark$                                                                                        & $\checkmark$                                                                                       & $\checkmark$                                                                               & $\times$                                                                                           & 20                                     \\ \hline
		File Sharing                               & $\times$                                                                                     & $\checkmark$                                                                                    & $\checkmark$                                                                                    & $\times$                                                                                     & $\times$                                                                                     & $\checkmark$                                                                                       & $\checkmark$                                                                                        & $\checkmark$                                                                                       & $\checkmark$                                                                               & $\times$                                                                                           & 47                                     \\ \hline
		Social Meida                               & $\checkmark$                                                                                    & $\times$                                                                                     & $\times$                                                                                     & $\times$                                                                                     & $\times$                                                                                     & $\times$                                                                                        & $\times$                                                                                         & $\times$                                                                                        & $\times$                                                                                & $\checkmark$                                                                                          & 147                                     \\ \hline
	\end{tabular}
\vspace{-.5em}
\end{table*}

$\bullet$ \emph{Diagnosis Summary.} For an expert, it requires dozens of iterations to provide in-depth analysis, resulting in extensive analysis records. Therefore, it is crucial to effectively summarize the key information from these records. To achieve this, we progressively summarize the lines of a record ($r_t$), which includes inputs for specific tools and the corresponding results, or relevant knowledge. Specifically, for each step $t$, we maintain a running summary ($s_{t-1}$), encapsulating previous actions and outcomes. Upon generating the new record $r_t$, an \llm is assigned to incorporate the main idea of $r_t$ into $s_{t-1}$, leading to the new summary, $s_t$. For a clearer explanation, consider the following example:
\begin{tcolorbox}[enhanced, sharp corners, width={8.5cm}, colback=white, boxsep=0mm, 
	borderline={.5mm}{0mm}{white, dashed}]
	{\bf [Current summary $s_{t-1}$]}\\
	- I know the start and end time of the anomaly.
	
	{\bf [New Record $r_t$]}\\
	Thought: Now that I have the start and end time of the anomaly, I need to diagnose the causes of the anomaly\\
	Action: is\_abnormal\_metric\\
	Action Input: \{``start\_time'': 1684600070, ``end\_time'': 1684600074, ``metric\_name'': ``cpu\_usage''\}\\
	Observation: ``The metric is abnormal''
	
	{\bf [New summary $s_t$]}\\
	- I know the start and end time of the anomaly.\\
	- {\it {\color{codegreen}I searched for is\_abnormal\_metric, and I now know that the CPU usage is abnormal.}}
\end{tcolorbox}

$\bullet$ \emph{Review Advice.} Next, each expert gives the improvement advice based on the diagnosis results and summarized procedures of other experts. The review prompt is written like \definecolor{deepgray}{gray}{0.3}
\textcolor{deepgray}{\bfit{$p_{review}$} = {\it ``Please review the above diagnosis results, and give necessary advice to correct the incorrect analysis or unclear results.''}}

$\bullet$ \emph{Diagnosis Refinement.} After the cross-review, each expert reevaluates their initial diagnosis and even conducts more inferences  (e.g., calling tools to analyze more relevant metrics or settings).  In this way, they can incorporate additional evidence, revising hypotheses, or overlooked aspects in the diagnosis results.

\bfit{Step5: Report Generation.} Based on both the refined diagnosis results, \emph{Expert Assigner} generate detailed diagnosis report for the given anomaly, including $(i)$ title (summary of the anomaly); $(ii)$ anomaly date; $(iii)$ detailed anomaly description (from alerts); $(iv)$ root causes (within diagnosis results); $(iv)$ solutions (within diagnosis results); $(v)$ summarized diagnosis process.

\section{EXPERIMENT RESULTS}
\label{sec:experiments}

With the carefully prepared micro benchmark, we conduct extensive experiments to evaluate the proposed techniques in \oursys 

\subsection{Environment Setup} 

\begin{sloppypar}
\hi{Database.} We implement \oursys in PostgreSQL 12.5, using $(i)$ the  pg\_stat\_statements plugin for tracking frequent queries, and $(ii)$ the hypopg plugin for creating hypothetical indexes~\cite{hypopg}. 
\end{sloppypar}

\hi{\llms.} We support prompt-based \llms including GPT-4-0613 and gpt-3.5-turbo-16k~\cite{openai}, where the {\it temperature} parameter is set to 0 in favor of reproduction. Fine-tuned \llms include \emph{Llama 2}, \emph{CodeLlama} and \emph{Baichuan 2}.

\begin{sloppypar}
\hi{Evaluated Methods.} The evaluated methods include: (1) \human. A human DBA with 2 years working experience analyze the root causes. 
(2) \dbotfour is the version of \oursys driven by GPT-4-0613 (within a limit of 8,192 tokens for each inference), which serves as 8 expert roles with different domain knowledge (Section~\ref{subsubsec: cluster}). 
(3) \dbotthree is the version of \oursys powered by the GPT-3.5 model. In case of exceeding the token limits, we use {\it gpt-3.5-turbo-16k} (a maximum of 16,385 tokens). 
(4) \dnn utilizes a shallow neural network (with two layers and ReLU activation) to classify the input abnormal metric vectors into one or multiple root causes~\cite{lianyuan2021ai}. 
(5) \decisiontree employs the decision tree algorithmto label the root causes for the input metric values~\cite{song2015decision}. 
(6) \vanillafour model that does not utilize the techniques in \oursys, which $(i)$ inputs suitable task description and demonstration examples and $(ii)$ outputs the root causes.  
(7) \vanillathree. Similarly, we test the performance of GPT-3.5 model without techniques in \oursys.
\end{sloppypar}


\hi{Ablation Methods.} \noindent We offer variants of \oursys for ablation analysis: (1) \noknow is \dbotfour that does not utilize the extracted knowledge. 
(2) \noreflect is \dbotfour that adopts the chain-of-thought reasoning (e.g., LangChain~\cite{langchain}). (3) \single is \dbotfour that utilizes single \llm to diagnose.

\subsection{Micro Diagnosis Benchmark}
\label{subsec:benchmark}

Based on works like~\cite{DBLP:journals/pacmmod/HuangWZTL023,liu2023opseval}, we design a micro benchmark that offers $(i)$ diversified anomaly scenarios (e.g., different applications, workloads, and anomaly types), $(ii)$ executable scripts, $(iii)$ clear scenario descriptions (e.g., {\it ``In a database of an e-commerce platform, 91 users simultaneously perform searches $\cdots$''}), together with $(iv)$ evaluation metrics that can reflect the diagnosis performance. 


\hi{Anomaly Cases.} As shown in Table~\ref{tab:micro-bench}, we include a diverse set of simulated applications: $(i)$ \textit{Internet of Things (IoT)} applications mainly have the ``highly commits'' anomalies, caused by handling a lot of incoming data from sensors; $(ii)$  \textit{E-commerce} applications exhibit multiple anomalies (e.g., ``highly updates'' and ``large data fetch''), possibly caused by concurrent updates to product databases and high volume data retrievals during sales; $(iii)$ \textit{Financial} applications involve anomalies like ``poor joins'', suggesting complex transactional operations; $(iv)$ \textit{Business Intelligence} applications mainly involve ``redundant index'' and ``missing index'' anomalies, emphasizing the importance of optimizing data access paths; $(v)$ \textit{File Sharing} applications (e.g., Dropbox, Google Drive) often encounter the ``large data fetch'' anomaly, caused by data retrievals of multimedia content. $(vi)$ \textit{Social Media} applications (e.g., MySQL originally for Twitter) predominantly face the ``highly commits'' anomaly when read and write data quickly. Different from {\it IoT}, they also involve complex queries that cause the ``correlated subquery'' anomaly.

\begin{sloppypar}
	\hi{Evaluation Metrics.} We adopt two metrics for practical diagnosis evaluation. First, similar to works like ~\cite{li2022actionable,DBLP:journals/pvldb/MaYZWZJHLLQLCP20}, we use \emph{Result Accuracy} ($\text{Acc}$) to quantify the precision of recommended root causes, i.e., 
	$$\text{Acc} = 
	\begin{cases}
	\frac{A_c - \sigma \cdot A_w}{A_a}, & \text{if } A_a > 0 \land A_c \geq \sigma \cdot A_w \\
	0, & \text{otherwise}
	\end{cases}$$
	where $A_c$ denotes the number of correct causes, $A_a$ denotes the total number of causes, $A_w$ denotes the number of wrongly detected causes, and $\sigma$ is a hyper-parameter with 0.1 as the default value, because we identify {\it redundant causes is less harmful than missing causes} and restrict to at most 4 root causes for an anomaly. 
\end{sloppypar}

Second, \emph{Human Evaluated Accuracy} ($\text{HEval}$) shares the same equation as $\text{Acc}$. However, $A_c'$ in $\text{HEval}$ denotes number of causes that $(i)$ are correctly detected and $(ii)$ the analysis process also makes sense (human evaluation). $\text{HEval}$ is {\it vital to provide reliable diagnosis for online usage}. 

\subsection{Performance Comparison}
\label{subsec:exp-nofinetune}

\begin{figure*}[!t]
	\centering
	\includegraphics[width=\linewidth]{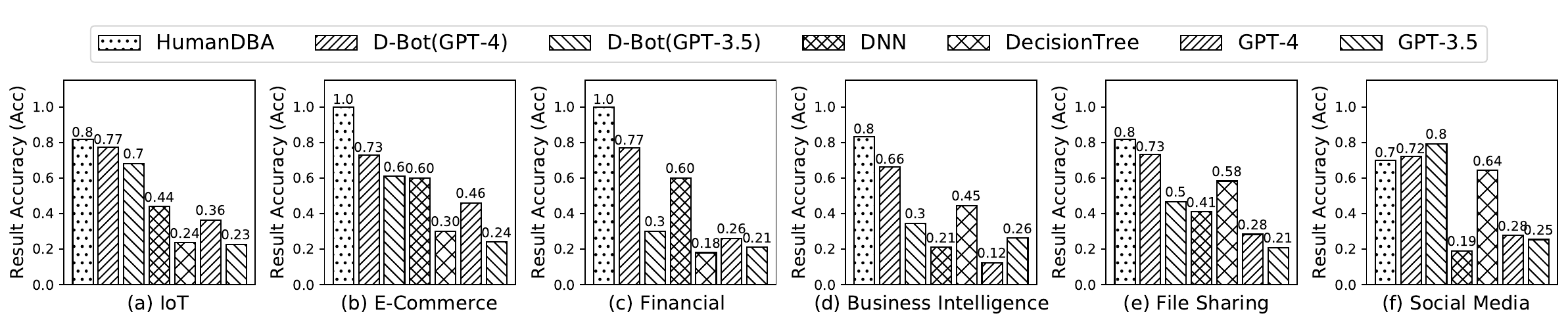}
	\vspace{-2em}
	\caption{{Performance Comparison (Result Accuracy). Note \emph{Acc} is a general numerical metric, which cannot fully reflect the diagnosis capacity (e.g., whether the diagnosis process is reasonable)}}
	\label{fig:acc}
	\vspace{-1em}
\end{figure*}

\begin{figure*}[!t]
	\centering
	\includegraphics[width=\linewidth]{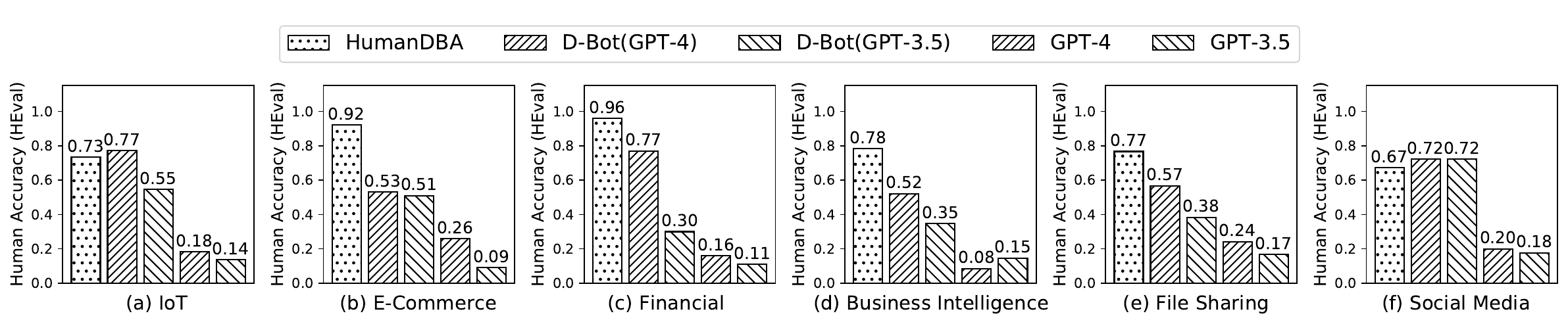}
	\vspace{-2em}
	\caption{{Performance Comparison (Human Evaluation). We do not include \dnn and \decisiontree because they either have the black-box problem or fail to provide root cause analysis that is easy to understand by humans.}}
	\label{fig:humaneval}
	\vspace{-1em}
\end{figure*}

We compare \oursys with three types of baselines, including manual diagnosis (HumanDBA),  existing machine learning methods (\dnn, \decisiontree), and origin \llms (GPT-4, GPT-3.5) across six applications. For each application, we sample ten testing anomalies from the micro benchmark. The remaining anomalies are used as the training samples for \dnn, \decisiontree. The performance results are illustrated in Figures~\ref{fig:acc}-\ref{fig:humaneval}.

\hi{Diagnosis Performance.}  \oursys achieves competitive performance as \human, such as outperforming \human with an accuracy of 80\% (\dbotthree) for the Social Media application. \oursys also demonstrates significant performance gains over the rest baselines (e.g., accuracy improvements ranging from 8\% to 54\% against \dnn and \decisiontree). The reasons are three-fold.

First, \oursys can judiciously utilize tools and provide informed diagnosis. 
For instance, It identifies specific problems such as \textit{``high memory usage due to heavy use of UPDATE and INSERT operations over the same tables''} by querying the \textit{pg\_stat\_statements} view. Conversely, the baselines struggle to detect the root causes, often defaulting to generic advice such as \textit{“resolve resource contention issues”}, which lack the specificity needed for actionable improvements, rendering them less effective in practical applications.

\begin{table}[!t]
	\small
	\caption{Performance on different anomalies. 
	}
	\vspace{-1em}
	\label{tab:your_label_here}	
	\centering
	\begin{tabular}{|c|p{1cm}|p{1cm}|p{1cm}|p{1cm}|}
		\hline
		\textbf{Diagnosis} & \multicolumn{2}{c|}{\textbf{\footnotesize{Single Cause Anomaly}}} & \multicolumn{2}{c|}{\textbf{\footnotesize{Multi-Cause Anomaly}}} \\
		\cline{2-5}
		\textbf{Method} & \textbf{\footnotesize{Acc}} & \textbf{\footnotesize{HEval}} & \textbf{\footnotesize{Acc}} & \textbf{\footnotesize{HEval}} \\
		\hline
		\human  & \underline{0.955} & \underline{0.720} & 0.487 & \underline{0.806} \\
		\hline
		\dbotfour & \underline{0.754} & \underline{0.500} & \underline{0.655} & \underline{0.669} \\
		\hline
		\dbotthree & 0.542 & 0.370 & \underline{0.533} & 0.493 \\
		\hline
		\dnn & 0.352 & N/A & 0.036 & N/A \\
		\hline
		\decisiontree & 0.331 & N/A & 0.086 & N/A \\
		\hline
		\vanillafour & 0.351 & 0.39 & 0.105 & 0.151 \\
		\hline
		\vanillathree & 0.266 & 0.2 & 0.144 & 0.130 \\
		\hline
	\end{tabular}
	\vspace{-1em}
\end{table}

\begin{figure*}[!t]
	\centering
	\includegraphics[width=\linewidth]{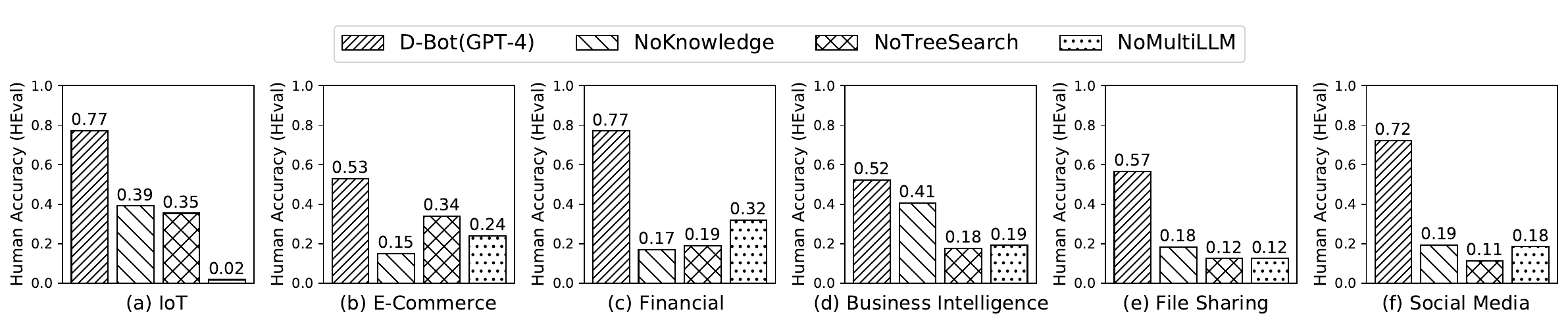}
	\vspace{-2em}
	\caption{{Ablation Study (Human Evaluation)}}
	\label{fig:ablation}
	\vspace{-.5em}
\end{figure*}

Second, \dbotfour owns contextual comprehension (\llm) and tree-search reasoning capability. For instance, with the reflection mechanism, \dbotfour can conduct comprehensive analysis and follow the most beneficial chain of actions (e.g., calculating the total cost of a plan and deciding the optimization actions).  In contrast, the baselines only input with basic abnormal metric values, perform general analysis, and often overlook underlying causes. For instance, in an \textit{INSERT\_LARGE\_DATA} case, \vanillafour merely identifies an increased count of running processes using the \textit{node\_procs\_running} metric, resulting in an early diagnosis termination. Moreover, \dnn and \decisiontree cannot leverage textual data, leading to their inability to resolve complex anomalies such as \textit{Poor Join}.

Third, \dbotfour utilizes the document knowledge to learn the analysis of potential performance bottlenecks like correlated-subquery structure. We find \vanillafour and \vanillathree tend to make unsupported hypotheses, leading to inaccurate diagnostics. For example, upon detecting \textit{SORT} operations in logged queries, \vanillathree inaccurately attributes the bottleneck to \textit{“frequent reading and sorting of large data volumes”}, missing query structure problems.  Compared to \human, \dbotfour is more careful in capturing important details that help find the root causes. For example, in \textit{Social Media}, \dbotfour does better than \human by collecting data from various sources (such as multiple system metrics and how query operators consume resources). This helps uncover problems like high I/O issues caused by concurrent inserts, which \human might ignore when focusing on a few slow queries.


\hi{Diagnosis Overhead.} {\it (1) Diagnosis Time.} \human needs one to two hours to write a diagnosis report even for typical anomalies.  This time is mainly consumed in devising solutions like indexing and query rewriting, even when the root cause is relatively straightforward. Instead, \oursys, takes ten to several minutes to diagnose relatively complex anomalies (e.g., {\it 5.38 minutes} for a composite anomaly $k$ with two root causes).  Because \oursys can efficiently interact with pre-equipped tools (context embedding) and enhance the efficiency (collaboration of multiple \llms). Traditional classifiers have lowest diagnosis time, as they simply map limited metrics to predefined causes. {\it (2) Diagnosis Expense.} Traditional classifiers and \oursys are more economical than \human. \dnn and \decisiontree require minimal system resources. And \oursys can save much manpower at a minimal financial cost (e.g., 1.8 dollar for diagnosing the anomaly $k$ with 40k \llm tokens).

\begin{tcolorbox}[enhanced,sharp corners,
	width={8.5cm},
	colback=white,
	borderline={0.3mm}{0.3mm}{white},
	boxsep=-.5mm]
	Finding 1. {\it \oursys achieves a remarkable improvement over baselines (8\% to 54\%) due to its advanced contextual understanding and knowledge and tool utilization, and even competes closely with human expertise.}
\end{tcolorbox}

\hi{Performance for Different Anomalies.} \dbotfour, while having lower accuracy in single cause anomalies (0.754), shows a remarkable consistency in multi-cause anomalies with an accuracy of 0.655. This consistency is also reflected in the \emph{HEval} scores (0.500 and 0.669, respectively), suggesting that \dbotfour maintains stable performance across different types of anomalies. \dbotthree and other methods like \dnn, \decisiontree, \vanillafour, and \vanillathree show a general trend of lower performance in both \emph{Acc} and \emph{HEval}, especially in multi-cause anomalies, highlighting the complexity of these scenarios that require advanced diagnosis methods like \oursys. Meanwhile, for \human, the \emph{HEval} scores are relatively high for both single (0.720) and multi-cause anomalies (0.806), demonstrating the necessity of understanding human experience.


\begin{tcolorbox}[enhanced,sharp corners,
	width={8.5cm},
	colback=white,
	borderline={0.3mm}{0.3mm}{white},
	boxsep=-.5mm]
	Finding 2. {\it \oursys provides a more balanced and reliable performance across diverse and complex anomaly types.} 
\end{tcolorbox}

\hi{\llm Factors.} The performance gap between \dbotfour and \dbotthree is significant, with \dbotfour outperforming \dbotthree by up to 30\% in accuracy and stability in applications. \dbotfour excels in generating precise tool calling commands and comprehensive diagnosis summaries. For instance, it adeptly identifies complex queries involving large table fetches, a task where \dbotthree often falls short. In contrast, \dbotthree is prone to producing more generalized and sometimes inaccurate action commands, leading to less effective outcomes.

\begin{tcolorbox}[enhanced,sharp corners,
	width={8.5cm},
	colback=white,
	borderline={0.3mm}{0.3mm}{white},
	boxsep=-.5mm]
	Finding 3. {\it A powerful \llm can benefit the  diagnosis performance of \oursys, which reflects in the effectiveness of prompt following and the depth of root cause analysis.} 
\end{tcolorbox}

\subsection{Ablation Study}

As shown in Figure~\ref{fig:ablation}, we verify the effectiveness of three main components in \oursys, i.e., document knowledge matching (\noknow), tree-search-based reasoning (\noreflect), and multi-agent diagnosis (\single).

\subsubsection{Document Knowledge Matching} Without the relevant knowledge in the prompt, \llm experts mainly rely on expert settings (i.e., role, task, steps) to call tools and analyze root causes. When comparing \noknow to \oursys, we observe a decrease in diagnosis accuracy ranging from 19.2\% to 64.1\%. We have two observations. First, \noknow produces significantly more redundant root causes (e.g., 2.05 times against \dbotfour), as it can't clearly tell apart relevant root causes using just the context. For instance, root causes like {\it ``many inserts''} and {\it ``large data insert''} both involve insert operations, but identifying them correctly requires specific knowledge about details like the number of insert operations and table sizes. Second, like the baselines, \noknow often provides very general diagnoses (e.g., ``abnormal patterns in CPU processes'') and fails to accurately identify many anomalies. Moreover, we also find that, although \llms like \vanillafour are pre-trained on open corpora, they need external knowledge matching (fine-tuning is limited in updating knowledge) for specialized tasks like database diagnosis.

\subsubsection{Tree Search Based Diagnosis.} \noreflect diagnoses less effectively than \dbotfour, showing a performance decrease by over 35.85\%.   It verifies that tree search plays an important role in correcting wrong knowledge matching or tool API callings (actions for extending child nodes), which significantly enhances the diagnosis accuracy, particularly for single-cause anomalies that involve various reasoning choices.  For instance, in scenarios such as identifying specific query-related issues or optimizing database knobs, tree search enables \dbotfour to navigate through multiple potential solutions and pinpoint the most effective one. 

\subsubsection{Multi-Agent Diagnosis.}  Our analysis verifies the effectiveness of multi-agent mode (\dbotfour) over single-agent mode (single). For instance, in the IoT application, \dbotfour achieves a 77.27\% success rate in identifying root causes, a substantial increase from the 39.09\% success rate of \single.  Besides, our tests on average diagnosis time revealed that \oursys (multi-agent mode) is more efficient compared to \single (single-agent mode). The reasons are two-fold. First, \oursys employs more than two experts in average (at most three), which utilize different metrics and domain knowledge to explore root causes and derive more root causes than \single. And these root causes are further examined, selected and refined during cross-review. Thus, \oursys achieves higher diagnosis accuracy than \single. Second, although \oursys takes time to select experts and conduct cross-reviews, the asynchronous mechanism reduces the iteration turns of tree-search algorithm in single experts, which generally take most diagnosis time. And so \oursys is also more efficient than \single in diagnosis time.


\begin{tcolorbox}[enhanced,sharp corners,
	width={8.5cm},
	colback=white,
	borderline={0.3mm}{0.3mm}{white},
	boxsep=-.5mm]
	Finding 4. {\it Techniques proposed in \oursys are crucial to boost diagnosis accuracy by reducing redundant root causes and enhancing precise anomaly identification.} 
\end{tcolorbox}





\begin{figure}[!t]
	\centering
	\includegraphics[width=1.05\linewidth]{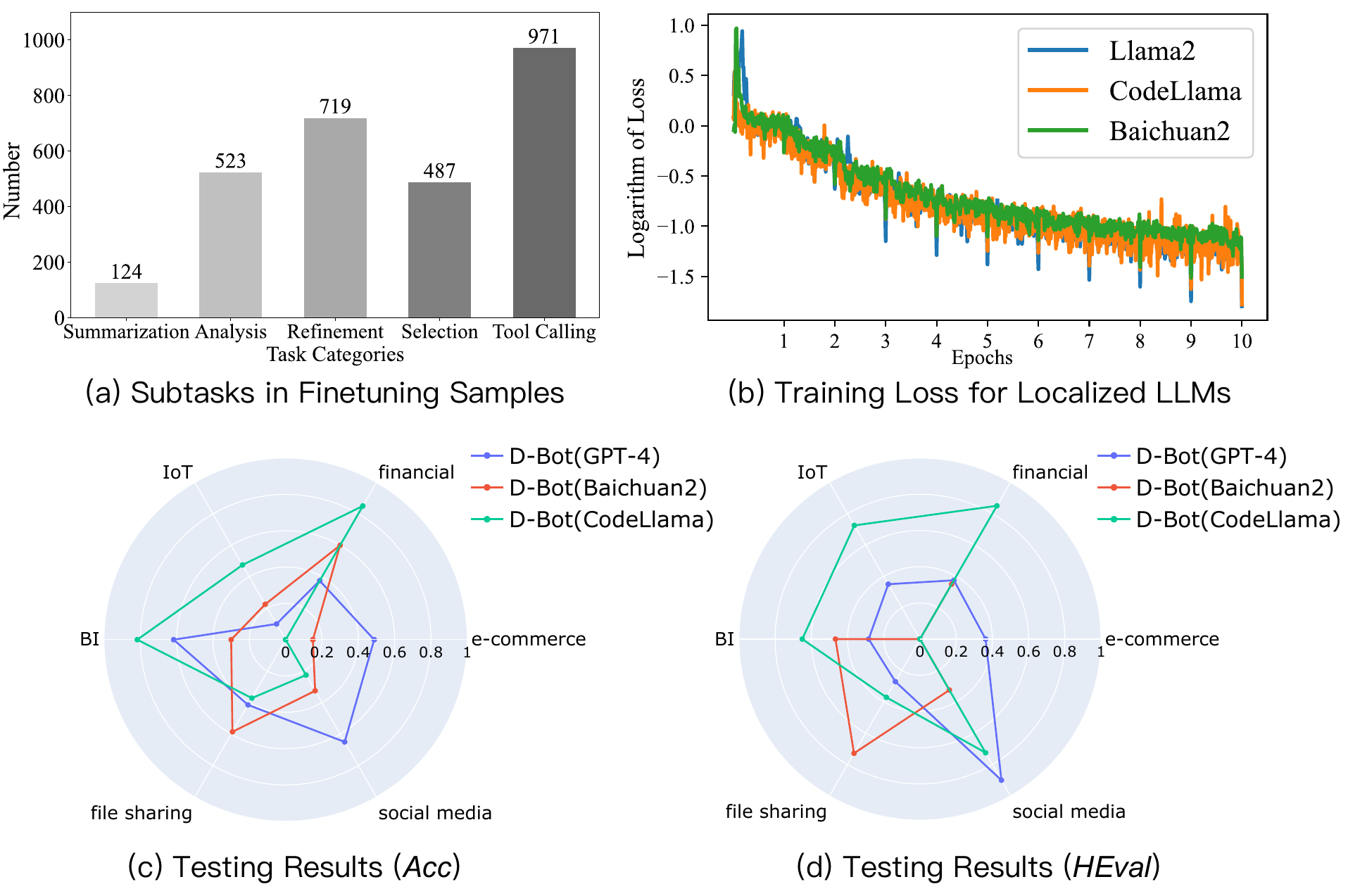}
	\vspace{-2em}
	\caption{Performance of Model Finetuning.}
	\label{fig:finetune}
	\vspace{-1em}
\end{figure}

\subsection{Model Fine-tuning}
\label{subsec:finetune}

\hi{Preparation.} We first record the diagnosis processes of \dbotfour consisting of 5 sub-tasks (e.g., tool calling) and 2819 samples in total (see Figure \ref{fig:finetune}(a)). We mix them together as a multi-task fine-tuning dataset. Specifically, the model input includes the prompt and historical messages, and we fine-tune \llms to simulate the corresponding \dbotfour response (after cleansed). \llms are implemented using PyTorch and BMTrain~\cite{zeng2023openbmb}, trained on a machine with 503 GB RAM and 1 NVIDIA A100 GPU.


\hi{Training Procedure.} We fine-tune three localized SOTA \llms, i.e., \emph{Llama 2-13B}, \emph{CodeLlama-13B}, \emph{Baichuan2-13B}. As shown in Figure~\ref{fig:finetune}(b), all \llms converge within 10 epochs. We then manually select the best epoch checkpoints (i.e., 4th epoch for \emph{Llama 2}, 1st epoch for \emph{CodeLlama}, 10th epoch for \emph{Baichuan 2}). Note {\it the obvious loss reduction does not mean increasing model performance}. We find many epoch checkpoints with low losses often over-fit the fine-tuning data (e.g., losing the text generation capabilities and tending to generate short confusing responses). Besides, \emph{Llama 2} cannot generate reasonable diagnosis results ({\it Acc} equals 0 for most cases) even in the best epoch. 

\hi{Performance Comparison.} As shown in Figure~\ref{fig:finetune}(c)-(d), the demonstrated \llms after fine-tuning achieve comparable performance to \emph{GPT-4} in 27 test cases. We have several observations. First, \emph{CodeLlama} performs best in financial, IoT and BI applications, because \emph{CodeLlama} is specialized for code generation, which is more sensitive to metrics and queries. For instance, it can accurately identify {\it slow queries involving multiple JOINs as root cause}.  
Second, \emph{Baichuan2} performs best for file application, which can assign suitable experts (e.g., Memory Expert), and analyze root causes in detail (e.g., pointing out disk I/O under-provisioned in the hardware configuration). However, the \emph{HEval} performance of \emph{Baichuan2} in financial application significantly degrades. For example, the model may list many root causes but does not give well-founded analysis. 
Third, \emph{GPT-4} performs best for e-commerce and media applications, and shows {\it balanced performance across all applications}. 
Moreover, the localized \llms show less generalizability to unfamiliar anomalies. For instance, the number of samples with {\it delete operations} as root causes is much smaller than others, causing the fine-tuned \llms to often fail in these cases.



\begin{tcolorbox}[enhanced,sharp corners,
	width={8.5cm},
	colback=white,
	borderline={0.3mm}{0.3mm}{white},
	boxsep=-.5mm]
	Finding 5. {\it \oursys using localized SOTA \llms can achieve comparable diagnosis performance to \dbotfour, but their generalizability is greatly affected by the fine-tuning samples.}
\end{tcolorbox}



\section{Related Work}



\hi{Database Diagnosis.} Existing works mainly rely on empirical rules and classification methods to analyze root causes. The ADDM tool~\cite{DBLP:conf/cidr/DiasRSVW05} maintains a graph of database resource modules, based on which they estimate the query execution time and infer the bottlenecks. DBSherlock~\cite{DBLP:conf/sigmod/YoonNM16} utilizes a decision-tree-like method to construct predicates (in the form of $Attr>k$). ISQUAD~\cite{DBLP:journals/pvldb/MaYZWZJHLLQLCP20} generates root causes by clustering queries with their metric vectors. However, these methods require great human intervention (e.g., designing rules,  features, labels). Besides, they lack some critical capabilities (e.g., accepting new contextual information, analyzing query logs) for real-world diagnosis. Although there are some \llm-based methods that incorporate maintenance knowledge~\cite{liu2023opseval}, they focus on general chatbot tools (e.g., Q\&A exercises) and also fail to conduct scenario-specific diagnosis.


\hi{\llm Agents.} Recent works have shown that \llms, when coupled with memory mechanisms and tools, can imitate human-like interactions and decision-making in a real-world context~\cite{DBLP:journals/corr/abs-2304-08354}. First, the augmentation of \llm agents with a variety of  tools --- ranging from web browser~\cite{DBLP:journals/corr/abs-2112-09332,DBLP:conf/acl/QinCJYLZLHDWXQL23} and wikipedia search~\cite{DBLP:conf/acl/TrivediBKS23,DBLP:conf/iclr/YaoZYDSN023}, to code interpreter~\cite{openinterpreter,DBLP:conf/icml/GaoMZ00YCN23,DBLP:journals/corr/abs-2211-12588} and multifaceted toolsets~\cite{DBLP:journals/corr/abs-2302-04761,DBLP:journals/corr/abs-2307-16789} --- has significantly enhanced \llm's adaptability. Besides individual agent skills, there is increasing interest in coordinating multiple \llm agents for collective intelligence~\cite{DBLP:journals/corr/abs-2304-03442,DBLP:journals/corr/abs-2308-08155,DBLP:journals/corr/abs-2303-17760,qian2023communicative,DBLP:journals/corr/abs-2305-14325,DBLP:journals/corr/abs-2308-00352}. Notably, AgentVerse~\cite{DBLP:journals/corr/abs-2308-10848} shows that teamwork among multiple \llm agents can perform better single agents in many tasks. \oursys presents an {\it \llm-powered diagnosis system in the multi-agent paradigm.}

\section{Conclusion}
\label{sec:conclusion}

 In this paper, we proposed a database diagnosis system leveraging large language models (\llms). We conducted offline knowledge extraction from documents and prepared function APIs from existing tools. We matched with suitable knowledge and APIs into \llm prompt for online diagnosis, and we proposed a tree search-based algorithm to accurately and effectively utilize tools and conduct analysis with knowledge. We designed a collaborative diagnosis mechanism that improved the efficiency with the collaboration of multiple \llms. Experimental results showed \oursys achieved remarkable improvements over baselines and human DBAs. 

\begin{acks}
We thank (1) Dongfeng Li for anomaly simulation support, (2) Wei Zhou for tool plugin assistant, (3) Xiaohui Nie, Dan Pei, Binyuan Hui, Chen Qian, Yu Shen for their valuable advice on this research.
\end{acks}

\clearpage
\balance
\bibliographystyle{plain}
\bibliography{tool}


\end{document}